\documentclass[a4paper,fleqn]{cas-dc}

\usepackage[numbers]{natbib}

\usepackage{amsmath,amssymb,amsfonts,upgreek,diagbox}
\usepackage{array}
\usepackage{algorithm}
\usepackage{algorithmic}
\usepackage{graphicx}
\usepackage{textcomp,makecell}
\usepackage{xcolor}
\usepackage{diagbox}
\usepackage{subfigure}
\usepackage{epstopdf}
\usepackage[justification=centering]{caption}
\usepackage{multirow}
\usepackage{makecell}
\usepackage{enumerate}
\usepackage{url}
\usepackage{color}
\usepackage{diagbox}
\usepackage{CJK}

\pdfoutput=1

\begin{document}
\bibliographystyle{unsrt}

\def\textpagefraction{.001}
\shorttitle{\textit{Gu et al. / Future Generation Computer Systems (2020)}}

\title [mode = title]{MSPPIR: Multi-Source Privacy-Preserving Image Retrieval in cloud computing}                      

\author[1]{Qi Gu}
\ead{634337549@qq.com}

\author[1]{Zhihua Xia}[orcid=0000-0001-6860-647X]
\ead{xia_zhihua@163.com}
\ead[URL]{http://www.mfsgroup.cn}
\cormark[1]

\author[1]{Xingming Sun}

\address[1]{Engineering Research Center of Digital Forensics, Ministry of Education, School of Computer and Software, Jiangsu Engineering Center of Network Monitoring, Jiangsu Collaborative Innovation Center on Atmospheric Environment and Equipment Technology, Nanjing University of Information Science \& Technology, Nanjing, 210044, China}
\cortext[cor1]{Corresponding author}

\begin{abstract}
Content-Based Image Retrieval (CBIR) techniques have been widely researched and in service with the help of cloud computing like Google Images. However, the images always contain rich sensitive information. In this case, the privacy protection become a big problem as the cloud always can't be fully trusted. Many privacy-preserving image retrieval schemes have been proposed, in which the image owner can upload the encrypted images to the cloud, and the owner himself or the authorized user can execute the secure retrieval with the help of cloud. Nevertheless, few existing researches notice the multi-source scene which is more practical. In this paper, we analyze the difficulties in Multi-Source Privacy-Preserving Image Retrieval (MSPPIR). Then we use the image in JPEG-format as the example, to propose a scheme called JES-MSIR, namely a novel \underline{J}PEG image \underline{E}ncryption \underline{S}cheme which is made for \underline{M}ulti-\underline{S}ource content-based \underline{I}mage \underline{R}etrieval. JES-MSIR can support the requirements of MSPPIR, including the constant-rounds secure retrieval from multiple sources and the union of multiple sources for better retrieval services. Experiment results and security analysis on the proposed scheme show its efficiency, security and accuracy.
\end{abstract}



\begin{keywords}
Searchable encryption \sep Privacy-preserving retrieval \sep Content-based image retrieval \sep multi-source
\end{keywords}

\maketitle

\section{Introduction}

Imaging device has rapidly become stronger and cheaper with the development of semiconductor technology. In this case, more and more high-resolution images are generated by people from all walks of life every day. The need for efficient storage and retrieval of images is more urgent by the increment of large-scale image databases among all kinds of areas. The development of cloud computing brings a suitable solution to the computation-intensive and storage-intensive image retrieval task, and many excellent image retrieval schemes \cite{zheng2017sift} have been proposed to put the CBIR into practical applications like Google Search By Image \cite{manning2008introduction}.

\begin{figure}[tb]
	\centering
	\includegraphics[width=1.0\linewidth]{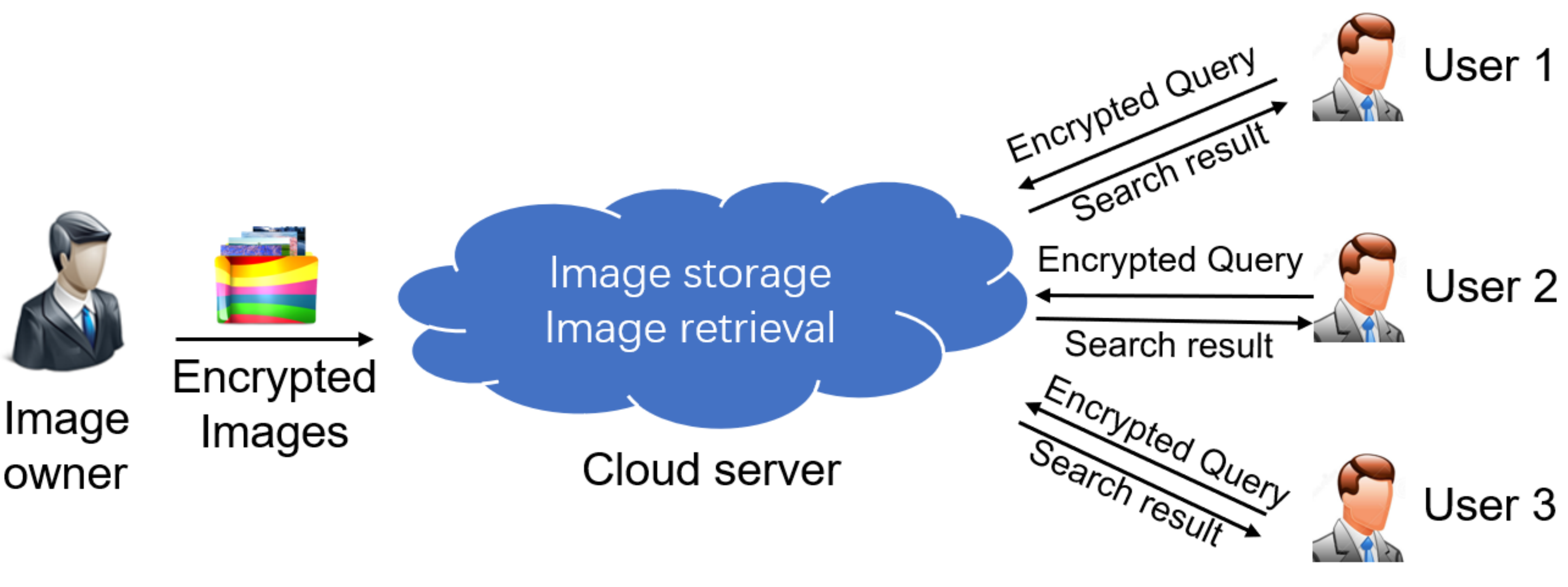}
	\caption{The system model with a single image owner.}
	\label{fig:commonsystemmodel}
\end{figure}

However, the images always contain rich sensitive information. What's more, in many cases, images are copyright restricted and the owners hope to profit from them by providing CBIR service. Therefore, it is unsafe to directly upload the unencrypted images to the cloud, which makes us drop into the dilemma between image retrieval and image security. Many prior works in the field of privacy-preserving CBIR (PPCBIR) have paid their attention to this problem. Briefly speaking, the image owner can upload the encrypted image features or the encrypted images to the Cloud Server (CS), and the CS can execute similarity computation between the encrypted data. A typical system model is shown in Fig. \ref{fig:commonsystemmodel}.

It should be noted that most of the existing schemes have a common limitation that they only consider the single-source (i.e., single image owner) case \cite{shen2018content}, where the image owner executes the authorizing and the authorized user retrieve the encrypted images of this owner with the help of CS. However, in real-world applications, image retrieval task is more likely to get multiple image sources involved. Firstly, the users of PPCBIR always hope that they can get more comprehensive search results. It is obvious that multi-source can cope with this problem better. Secondly, the image owners can enhance their competitiveness by uniting and providing their services to the authorized users together. Last but not least, the CS is more willing to service for multi-source to enhance stability and profitability as they can provide more computation and storage services. The joint demand of all entities makes multi-source an indispensable choice. Some recent works \cite{shen2018content, zhang2017pic, ferreira2017practical} have noticed the significant meaning of these scenes, however, to the best of our knowledge, no existing paper comprehensively considers MSPPIR and gives the scheme safely and efficiently.

The introduction of MSPPIR will lead to two new problems. The first one is that how can the user retrieval from different image owners at the same time, in other words, how to compare the distances between encrypted images which are encrypted by different keys. The second one is how to let the a part of image owners can provide for the user together, which means the user can use the same key to retrieve from a union of image owners. It is clear that CBIR is a real-time task. However, to ensure the image security, especially in the multi-source scene, is quite a challenge to efficiency. Besides, as the image encrypted in the spatial domain cannot be compressed a lot, two PPCBIR works \cite{cheng2016encrypted, zhang2014histogram} try to encrypt images in JPEG-domain. However, these schemes still suffer from the problems like feature leakage, index lacking, etc.

To address these challenges, we propose a new secure scheme JES-MSIR for MSPPIR, in which we consider two basic requirements that are different from the scenario with a single image owner.  In total, the contributions of this paper can be summarized as follows:

\begin{itemize}
	\item[1)] We formally define the MSPPIR problem in terms of functionality and security. Firstly, the authorized user should be able to execute the retrieval from all the owners, who authorize to him, with constant (i.e., irrelevant to the number of owners) rounds of communication to the CS. Secondly, a part of the owners should be able to unite as a group to provide the retrieval service together. Finally, the security should be considered under the reasonable threat model.
	
	\item[2)] We propose a novel scheme which can support MSPPIR efficiently. The permutations are used to ensure the security and accuracy. The property of permutation is further exploited to deal with the collusion problem and support the union of sources. The image encrypted is conducted with the quantized DCT coefficients in JPEG-format images to avoid the file expansion. The bag-of-words (BOW) and multiple permutations are utilized to cope with the problems like low retrieval efficiency and feature leakage in the existing JPEG-domain single-source PPCBIR schemes.
	
	\item[3)] We make detailed experiments on two real-world image databases. It is shown that the efficiency and retrieval accuracy of our scheme is better than the existing schemes which just partly support multi-source, and the security is on par with the existing PPCBIR schemes.
	
\end{itemize}

The rest of our paper is organized as follows. Section \ref{sec:section2} summaries the related works, especially, we give explanations about why most of the single-source schemes are not suitable for the multi-source scene. Section \ref{sec:section3} introduces the system architecture and preliminaries. The detailed scheme design is presented in Section \ref{sec:section4} and Section \ref{sec:section5}. The Section \ref{sec:section6} gives the security analysis. Experiment results are shown in Section \ref{sec:section7}. Finally, conclusions are made in Section \ref{sec:section8}.

\section{Related work}\label{sec:section2}

Existing schemes on PPCBIR can be briefly classified into two categories. In the first category, the image owner firstly extracts the aggregated feature from plaintext image, then use specific encryption methods to encrypt the feature or index. The image owner uploads the encrypted features and encrypted images to CS at last. CS can execute the retrieval in the encrypted domain with specific similarity measurement methods. In the second category, the owner only needs to encrypt the image, the tasks of feature extraction and index building are all undertaken by CS, which makes an ideal environment for MSPPIR. The kernel difference is that feature extracted before or after the encrypted image upload and we detailedly discuss these schemes in the following.

\subsection{Feature-encryption based schemes}

Since the feature extraction task is undertaken by the owner, the kernel task of schemes in the first category is constructing a functional encryption on the feature to make the distance between encrypted features valid. The methods can be broadly divided into two classes \cite{lu2014confidentiality}: those based on randomization symmetric encryption techniques and those based on homomorphic encryption. To our knowledge, Lu \textit{et al.} \cite{lu2009enabling} proposed the first PPCBIR scheme over the encrypted image database. This scheme uses the min-hash algorithm and order-preserving encryption to protect the visual words which are utilized to represent the images. In another work, Lu \textit{et al.} \cite{lu2009secure} investigates three image feature protection techniques including bit-plane randomization, random projection, and randomized unary encoding. Based on the property of bit computation, the encrypted feature is still valid for retrieval. Xia \textit{et al.} \cite{xia2015towards} proposed a PPCBIR scheme based on Scale-Invariant Feature Transformation (SIFT) \cite{lowe1999object} features and Earth Mover's Distance (EMD) \cite{rubner2000earth}. The calculation of EMD is a linear program problem, and a linear transformation was utilized to protect the privacy information during the solution process of the EMD problem.The above methods all belong to the randomization symmetric encryption techniques.

Homomorphic encryption (HE) technology is a cryptography technology based on the assumption of computational intractability. Some early works \cite{shashank2008private,zheng2013efficient} considered the secure distance computation of feature vector based on the Somewhat Homomorphic Encryption (SHE) \cite{paillier1999public} which can support addition or multiplication on ciphertext, however, they are not a practical scheme in the PPCBIR as they expose part of plaintext feature. To the best of our knowledge, Lu \textit{et al.} \cite{lu2014confidentiality} firstly pointed out that the SHE methods can not complete the secure retrieval without the interactions with the authorized user. They further prove that although CS can execute the retrieval based on Fully Homomorphic Encryption (FHE) \cite{gentry2009fully} technology which can support the addition and multiplication on ciphertext, the time and storage consumption is far more than the methods based on randomization distance-preserving encryption. The other schemes \cite{shen2018content,zhang2017pic} in this type will be detailed described later.

\subsection{Image-encryption based schemes}
The strategies in the first category suffer from a common disadvantage. As the big volume of storage and large computation complexity, both the image feature extraction and index construction are resource-consuming operations. In this case, the researchers try to outsource the feature extraction task to the cloud, which brings up the methods in the second category. Similar to the first category, the methods in the second category can be briefly classified into two classes. The first one tries to extract the encrypted classic feature (e.g., SIFT) from the encrypted images through SHE technology, and the second uses invariant statistics as the feature based on random encryption. To our knowledge, Hsu \textit{et al.} \cite{hsu2012image} was the first to investigate privacy-preserving SIFT in the encrypted domain by utilizing the Paillier cryptosystem. However, their scheme is computationally intractable and insecure \cite{schneider2014notes}. The following schemes \cite{hu2016securing,wang2016catch} in this class try to improve their practicability by using two CS work collaboratively. In recent years, more researchers \cite{liu2019intelligent} try to use pre-trained VGG16 as the feature extractor to extract encrypted features. However, the time and storage consumption taken by HE on image and plenty of interactions between servers is still hard to accept. The essential reason for high complexity is the large number of nonlinear feature extraction operations on the encrypted images.

The methods based on statistics is the scheme where users and CS are both low computational cost, and it makes these schemes become the most practical one. Ferreira \textit{et al.} \cite{ferreira2017practical} proposed a tailor-made Image Encryption Scheme called IES-CBIR. In this scheme, the random permutation is employed to protect the value (i.e., color) information and the position (i.e., texture) of image pixels is shuffled randomly. After encryption, the owner sends the encrypted images to CS. The encrypted HSV (Hue-Saturation-Value) color histograms will be further extracted at the cloud server side. The Hamming distances on these histograms are finally used to evaluate the similarities between the corresponding images. The global histogram is undesirable for CBIR, therefore, Xia \textit{et al.} \cite{xia2019boew} further extracted local histograms as the local features, and get the aggregating feature with the help of BOW. However, the schemes in the spatial domain will destroy the image compression, and the encrypted images have to be stored in lossless-compression format (PNG, zip), thus it will bring extra storage and time consumption. A valid solution to this problem is to encrypt the image in JPEG-domain and keep the JPEG-format be hold after the encryption. Zhang \textit{et al.} \cite{zhang2014histogram} encrypts the JPEG image by permuting the DCT coefficients of different blocks at the same frequency position, and Cheng \textit{et al.} \cite{cheng2016encrypted} permutes the entropy-coded segments in the JPEG bitstream. However, these schemes exposed the feature of plaintext. To cope with this problem, Liang \textit{et al.} \cite{liang2019huffman} encrypted the Huffman-code histograms. However, their scheme is still fragile in the Known-Background-Attack (KBA) model. What's more, the previous works in JPEG-domain did not give the feature aggregation scheme, which will make the retrieval time unacceptable.

\subsection{Partly supported MSPPIR}
Although most of the mainstream schemes in the PPCBIR are mentioned above, few papers above considered the scene of multi-source. A straightforward idea is extending existing schemes to the multi-source scenario by executing searching over encrypted images belong to different owners one by one. However, it will introduce plenty of rounds of communications between authorized users and CS. A necessary improvement is performing multiple retrievals at constant rounds (e.g., one interaction). However, in this case, the randomization based scheme in the first category is vulnerable to the malicious image owner. For example, the stream cipher key is exposed when the attacker gets the ciphertext and its plaintext, which makes the collusion between the image owner and CS become a threaten. Besides, the union expansion of schemes in the first category is still an open problem. The scheme based on classic feature extraction in the second category is also unsuitable for the multi-source expansion as the time-consumption will be more unacceptable. Some other methods unmentioned above like partial encryption \cite{gong2018privacy,xu2017privacy} are also unsuitable for the scalability in that the security risk in the single-source will be more magnified.

To our knowledge, there are three existing schemes which partly support the multi-source scene. Shen \textit{et al.} \cite{shen2018content} firstly point out the significant meaning of retrieval multi-source in one interaction, they propose a scheme called MIPP based on the methods in secure multi-party computation (SMC) \cite{jung2014collusion}, which supports the sum of ciphertext is same as the sum of plaintext. The scheme lets each image owner encrypt their feature vector by their own key and use their sum as an evaluation of distance. To avoid the interaction between the image owner and authorized user during the retrieval, a key management center (KMC) is introduced to decrypt the image belong to the image owner, then encrypt it with the key from the authorized user. However, on the one hand, this evaluation is not suitable for the image, which makes their retrieval accuracy not good. On the other hand, the scheme exposes the sum of plaintext features to the CS and exposes the plaintext image to the KMC, which makes their scheme insecure. Zhang \textit{et al.} \cite{zhang2017pic} proposed a feasible scheme called PIC based on multi-level FHE which supports the key conversion in the encrypted domain \cite{xiao2012efficient}. The CS and KMC both possess a part of the secret key. When a user adds into the system, the trusted party (TP) distribute the secret key to the user, CS, and KMC, which makes the ciphertext can be transformed to the same key by the collaborative computing. Although their scheme can get similar accuracy with the plaintext, the time consumption caused by multi-level FHE is unacceptable. Besides, the security of this scheme is based on a global secret key, which make it vulnerable to the collusion attack (i.e., the collusion between the CS and KMC). The above two schemes can be regarded as the scheme in the first category and they only consider the scene about one single user authorized by multi-source. In \cite{ferreira2017practical}, a brief discussion about the union between the owners is given. In their scheme, an owner creates the repository, when the other users join in, they need to use the repository key to encrypt the pixel color features, and the users can encrypt their pixel positions on their own. However, it will make the image owner execute extra consumption when they join in a repository. What's more, the following works in the second category are all paying little attention to the scene that users authorized by multi-source.

As the description above, existing schemes on MSPPIR are suffering from the shortage of accuracy, security, efficiency, and scalability. Inspired by existing schemes based on invariant statistics in the second category, we propose a novel system model in Section \ref{sec:section3}, and show the complete scheme in Section \ref{sec:section4} and Section \ref{sec:section5} to cope with challenges on multi-source PPCBIR scene.

\section{System model and preliminaries}\label{sec:section3}
\unskip
\subsection{System model}\label{sec:systemmodel}

\begin{figure}[tb]
	\centering
	\includegraphics[width=1.0\linewidth]{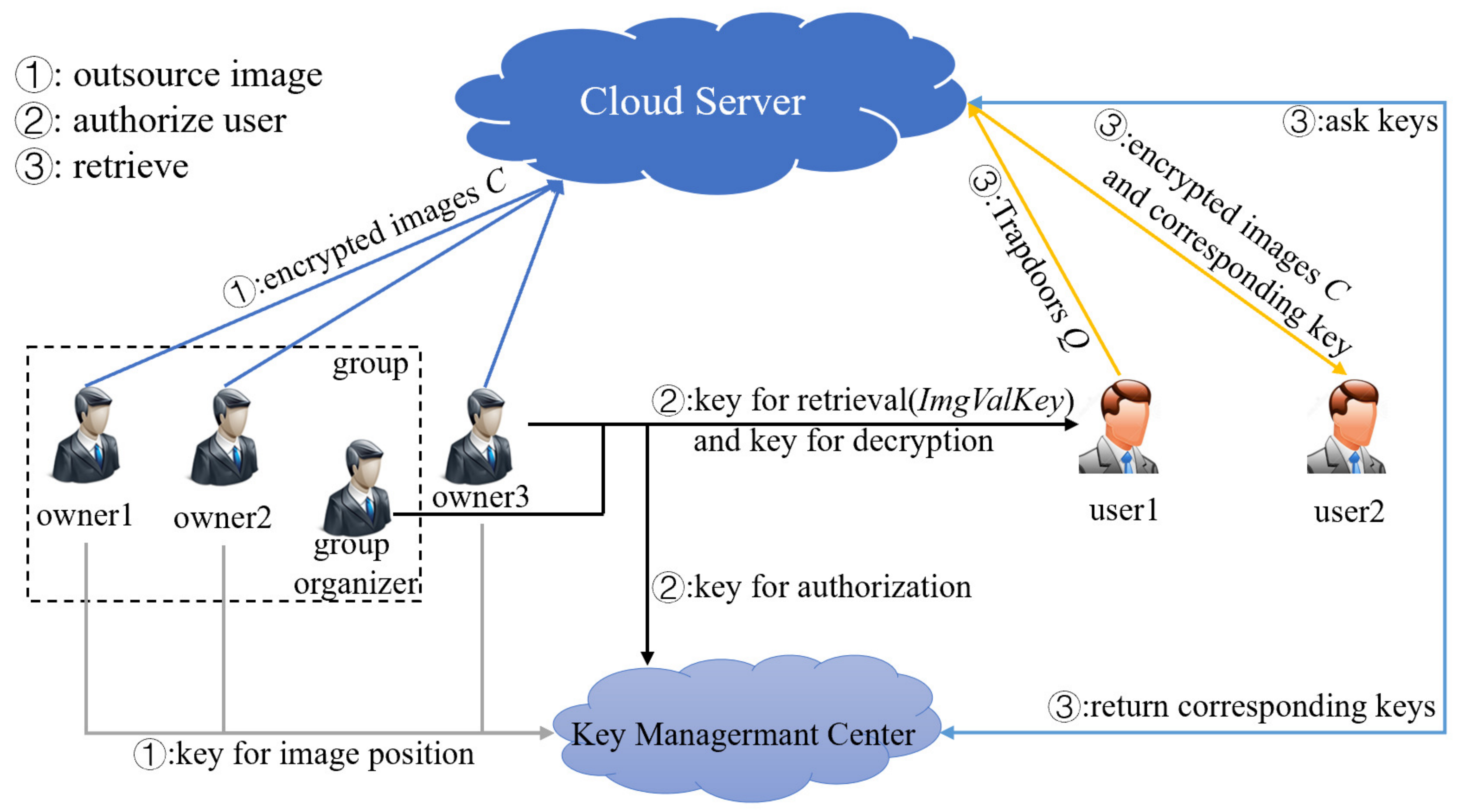}
	\caption{System model. In the figure, the processes of outsourcing images, authorizing users, and retrieval are shown.}
	\label{fig:systemmodel}
\end{figure}

Similar to \cite{zhang2017pic} and \cite{ferreira2017practical}, the proposed system involves five types of entities, i.e., the image owner, the group, Cloud Server (CS), Key Management Center (KMC), and the Authorized User (AU), as shown in Fig. \ref{fig:systemmodel}.

\textbf{\boldmath{Image owner}} has a corresponding identity which can be called as $OID$. Each image owner has a large-scale image database $~\mathcal{I}_{OID} = \left\{ I_{i} \right\}_{i = 1}^{n_{OID}}$. The images belong to an owner have a corresponding identity set $\mathcal{IID}_{OID} = \left\{ IID_i \right\}_{i=1}^{n_{OID}}$, where the $n_{OID}$ means the number of the images the owner has. To preserve the security of images, before uploading, each owner generates an encrypted image set $~\mathcal{C} = \left\{C_i \right\}_{i=1}^{n_{OID}}$ from $~\mathcal{I}_{OID}$ by image encryption.

\textbf{\boldmath{Group}} with a group identity $GID$ is a union of several image owners. The goal of group is to give better service for the authorized users, which means after authorization by the group, the user can search all the images belong to the image owners in the group without interaction with them. The \textbf{\boldmath{group organizer}} takes the responsibility to authorize the users and update the members. The group organizer should be trusty to all members of the group. The group organizer could be undertaken by a trusty member or trusty third party.

\textbf{\boldmath{Cloud Server}} stores the encrypted images from the image owners and provides CBIR service for users. What'more, the cloud server will further extract aggregate features from the encrypted images for the owner and group to get better retrieval time.

\textbf{\boldmath{Key Management Center}} takes the responsibility for storing the key for each image and the authorized information. When an owner adds a new encrypted image to the CS or authorizes a user, he will send a corresponding key to the key management center. Two keys will be got when the group organizer authorizes a new user. During the retrieval, CS needs one interaction with KMC.

\textbf{\boldmath{Authorized User}} with a corresponding identity called as $UID$ can be authorized by multiple owners/groups, and they can get the retrieval results from all the authoring sources with a single interaction. Authorized users need no interactions with KMC.


\subsection{Security model}\label{sec:securitymodel}

Similar to previous schemes \cite{zhang2017pic,ferreira2017practical}, the honest-but-curious CS and KMC are considered in our scheme, i.e., they will follow the protocol specification, but may try their best to harvest the content of the encrypted images. In general, CS and KMC are well protected, so we don't consider compromise attack in this paper.

As we could set the CS and KMC in different service providers (i.e., Google and Amazon) and the authorized user will not interact with the KMC, we assume that it is not possible to have an authorized user who colludes with both CS and KMC. Similar, as the union of group implies the assumption that owners in the group is highly believable, we assume both CS and KMC collude with an image owner in one group is impossible. As the group organizer need not do more thing besides key generation and distribution, we assume the group organizer is trusty to all the members in the group. Same as previous schemes \cite{zhang2017pic}, the collusion between entities only include the existing information exchange, the further defraud collaboratively is beyond consideration as it is easy to be detected by the other non-collusion server. Please note that the security model in this paper is stricter than all previous works related to MSPPIR \cite{shen2018content,zhang2017pic,ferreira2017practical}.

\subsection{Preliminaries}\label{sec:preliminaries}

\subsubsection{Overview of JPEG encoding}\label{sec:JPEGencoding}

More than $95\%$ of the images in web \cite{schaefer2017fast} is JPEG-format. Generally speaking, it will be faster to operate the JPEG image without decoding. To better explain the image encryption and feature extraction operations, we here briefly introduce color JPEG encoding.

As we all know, the color image is composed of a number of pixels that are represented by RGB values. Based on the characteristics that human eyes are insensitive to chrominance information and high-frequency information, the JPEG encoding firstly transforms the RGB information to YUV pattern, then down-sampling the UV information (e.g., Y:U:V = 4:1:1). Then the image is split to a series of 8$\times$8 non-overlapped blocks, and DCT transformation is executed on each block. As a result, the RGB values in a block are transformed as one DC value and 63 AC values. At last, the quantization table is used to compress these values, the higher frequency information which means the relatively later part of AC values will be strongly squeezed, and most AC values will be squeezed to zero.

Due to the dependency of the adjacent image blocks, the difference value between two sequential DCT blocks is calculated to represent the DC value. As most of the AC values are zero, the zig-zag scan and run-length encoding are used to encode 63 AC values in each block. For example, suppose the zig-zag sequence of a block is $\{ $3,-8,0,-1,0,0,0,3,0,0,-4,EOB$\}$, it can be converted into several $(r,v)$ pairs: $\{$(0,3), (0,-8), (1,-1), (3,3), (2,-4), (0,0)$\}$, where $r$ denotes the number of zeros before a non-zero AC coefficient whose value equals to $v$. The symbol EOB (End-Of-Block) implies that all remaining AC coefficients in the block are zero, and denotes as one specific pair (0,0). The Huffman code and VLI code table are finally used to encode the DC difference value and $(r,v)$ pairs with the VLI code table shown in Table \ref{tab:VLIcode}, one VLI code is composed by the bitstream and the group it belongs to. For simplicity, in the following, the bitstream of DC difference value will be called $DC$, and the group index of $DC$ will be denoted as $g_{DC}$. It is easy to find that the $DC$ and $(r,v)$ pairs contain nearly all the information of the images.

\begin{table*}[!t]
	\centering
	\renewcommand{\arraystretch}{1.3}
	\caption{Variable-Length Integer (VLI) coding table}
	\label{tab:VLIcode}
	\begin{tabular}{|c|c|c|}
		\hline
		Value $v $  & {Group Index (Number of Bits)} & {Binary Code}\\
		\hline
		0 & 0 & -\\
		\hline
		-1, 1 & 1 & 0,1\\
		\hline
		-3, -2, 2, 3 & 2 & 00, 01, 10, 11\\
		\hline
		-7, -6, -5, -4, 4, 5, 6, 7 & 3 & 000,001,010,011,100,101,110,111\\
		\hline
		-15,\dotso ,-8, 8,\dotso ,15 & 4 & 0000, 0001,\dotso ,1110,1111\\
		\hline
		-31,\dotso ,-16, 16, 31 & 5 & 000000,\dotso ,111111\\
		\hline
		-63,\dotso ,-32, 32,\dotso ,63 & 6 & 0000000,\dotso ,1111111\\
		\hline
		-127,\dotso , -64, 64,\dotso , 127 & 7 & 00000000,\dotso ,11111111\\
		\hline
		-255,\dotso , -128, 128,\dotso ,255 & 8 & 00000000,\dotso ,111111111\\
		\hline
		-511,\dotso ,-129,129,\dotso ,511 & 9 & 000000000,\dotso ,1111111111\\
		\hline
		-1023,\dotso ,-512,512,\dotso ,1023 & 10 & 0000000000,\dotso ,11111111111\\
		\hline
		\dotso & \dotso & \dotso \\
		\hline
	\end{tabular}
\end{table*}

\subsubsection{Bag-of-word model}

CBIR technologies extract visual features to represent the images. In the early stages of its development, global features \cite{jain1996image} are extracted from the image to perform the retrieval. However, the global feature is always easy to be affected by the illumination and rotation, etc. The local features (e.g., SIFT) are used to cope with this problem. However, the local features are always too large and unstable, in this case, the feature aggregation methods are gradually developed. The BOW (Bag-Of-Word) \cite{sivic2003video} is one of the most popular models. There are three steps in the BOW model:

\emph{(i) Local histogram extraction}. The first step is to extract local features from the images in the database. The local feature(e.g. SIFT) is commonly used in the CBIR. However, to encrypted images, on the one hand, without suitable aggregation methods, the local feature is not outstanding in the low-resolution images; on the other hand, the non-linear detection schemes which local feature uses make the feature extraction from encrypted image difficult \cite{hsu2012image,hu2016securing}.

\emph{(ii) Vocabulary generation}.  The second step is to construct the visual vocabulary. Typically, $k$-means method can be employed to cluster the local features into $k$ classes. The cluster centers are defined as visual words. The full set of visual words constitute the vocabulary.

\emph{(iii) Histogram calculation}. The last step is to calculate the histogram of visual words. All the local features are represented by their nearest visual words. Finally, each image is represented by a $k$-bins histogram of visual words. It should be noted that the position of the visual words has been ignored in this way, and it gives the space for the encryption.

\subsubsection{Permutation and Bitxor}

The permutation encryption and bitxor encryption is widely used in the encryption on features \cite{lu2009enabling} and images \cite{ferreira2017practical,cheng2016encrypted,xia2019boew}. In these schemes, symmetric secret keys are used during the encryption and decryption. The permutation-based encryption and decryption is presented in algorithm \ref{alg: encperm} and algorithm \ref{alg: decperm}.

\begin{algorithm}[t]
	\caption{EncPerm}
	\label{alg: encperm}
	\begin{algorithmic}[1]
		\REQUIRE
		Plaintext Data $D = \left(d_1, d_2,\dots, d_n \right)$, Permutation key $K = \left(k_1, k_2, \dots,k_n \right)$
		\ENSURE
		Encrypted Sequence $C = \left(c_1, c_2, \dots, c_n \right)$\\
		$//$ $d_i,k_i,c_i \in \left\{1,2,\dots,n \right\}$
		\FOR{each $i \in E_n$}
		\STATE $c_i = d_{k_i}$
		\ENDFOR
	\end{algorithmic}
\end{algorithm}

\begin{algorithm}[t]
	\caption{DecPerm}
	\label{alg: decperm}
	\begin{algorithmic}[1]
		\REQUIRE
		Encrypted Data $C = \left(c_1, c_2,\dots, c_n \right)$, Decryption key $K = \left(k_1, k_2, \dots,k_n \right)$
		\ENSURE
		Plaintext Data $D = \left(d_1, d_2, \dots, d_n \right)$\\
		$//$ $d_i,k_i,c_i \in \left\{1,2,\dots,n \right\}$
		\FOR{each $i \in E_n$}
		\STATE $d_{k_i} = c_{i}$
		\ENDFOR
		
	\end{algorithmic}
\end{algorithm}

For simplicity, the orderly sequence of positive integers from 1 to $N$ is denoted as $E_N$. For example, $E_3$ denotes the sequence $(1,2,3)$. In the existing schemes \cite{ferreira2017practical,cheng2016encrypted,lu2009secure,xia2019boew}, the permutation key is used to encrypt the plaintext features or images, which means input data of \textsf{EncPerm} can be seen as $E$. To meet the needs of MSPPIR, the transformation between secret keys are further considered. From the basic properties of the \emph{permutation group} \cite{Permuta}, it is easy to get formula \ref{eq:decencenc}.

\begin{equation}
\label{eq:decencenc}
\begin{array}{c}
\textsf{EncPerm}(\textsf{DecPerm}(K_2,K_1),\textsf{EncPerm}(K_1,K)) \\
= \textsf{EncPerm}(K_2,K)

\end{array}
\end{equation}

Specially, we can get formula \ref{eq:encenc} when we set $K = E$.

\begin{equation}
\label{eq:encenc}
\begin{array}{c}
\textsf{EncPerm}(\textsf{DecPerm}(K_2,K_1),\textsf{EncPerm}(K_1,E))\\
= \textsf{EncPerm}(K_2,E)

\end{array}
\end{equation}

It should be noticed that for each $D$, there is a corresponding $K$ can get the same $\textsf{DecPerm}(D, K)$. The primary fact means that it will be difficult to infer $D$ or $K$ from the $\textsf{DecPerm}(D, K)$ only. Based on same reason, it further implies that exposed $\textsf{DecPerm}(K_2, K_1)$ and $\textsf{EncPerm}(K_1, K)$ will not leak the $K$, $K_1$ and $K_2$. It is easy to notice that the BitXor (Bit-wise XOR) computation has the same property.

\subsubsection{Notations}

\begin{itemize}[]
	\item \textbf{$\mathcal{K}_p$}, \textbf{$\mathcal{K}_v$}, \textbf{$\mathcal{K}_u$}: secret keys for generating $ImgPosKey$, $ImgValKey$, $UserKey$.
	\item \textbf{$ImgPosKey$}, \textbf{$ImgValKey$}: secret keys for protecting image position, value.
	\item \textbf{$UserKey$}: secret keys for protecting all the potential $ImgPosKey$.
	\item \textbf{$key_{blo}$}, \textbf{$key_{inblo}$}, \textbf{$key_{dc}$}: secret keys in $\mathcal{K}_p$ which are used for generating $pmtb$, $pmtp$, $bitdc$.
	\item \textbf{$pmtb$}, \textbf{$pmtp$}, \textbf{$bitdc$}: secret keys for protecting inter-block information, intra-block information, bit information of $DC$ and $v$.
	\item \textbf{$key_v$}, \textbf{$key_l$}: secret keys in $\mathcal{K}_v$ which are used for generating $pmtv$, $pmtDCL$.
	\item \textbf{$pmtv$}, \textbf{$pmtDCL$}: secret keys for protecting bit-length information of $v$ and $DC$.
	\item \textbf{$key_{Ublo}$}, \textbf{$key_{Uinblo}$}, \textbf{$key_{Udc}$}: secret keys in $\mathcal{K}_u$ which are used for generating $Upmtb$, $Upmtp$, $Ubitdc$.
	\item \textbf{$Upmtb$}, \textbf{$Upmtp$}, \textbf{$Ubitdc$}: secret keys for protecting all potential $pmtb$, $pmtp$, $bitdc$.
	\item \textbf{$ImgPosKey'$}, \textbf{$EncImgPosKey'$}: secret format of the $ImgPosKey$ stored in KMC, secret format of the $ImgPosKey$ used for decryption.
	\item \textbf{$IncUsrKey$}, \textbf{$IncValKey$}: secret keys for adjusting $UserKey$, $ImgValKey$ during joining a group.
	
\end{itemize}

\section{Basic scheme}\label{sec:section4}
\unskip

In this section, we only consider the scene that the authorized user search from multi-owners, the enhancement on security and the scheme of group union will be given in the next section. The proposed scheme is given from the perspective of different entities.

\subsection{Owner Side}\label{sec:ownerside}

\subsubsection{Image Key Generation}

As mentioned in subsection \ref{sec:JPEGencoding}, the JPEG-format image is mainly made up of $DC$ values and $(r,v)$ pairs. Similar to the image in spatial-domain, the image can be separated into two kinds of information, i.e., value information and position information. To protect the image content, we firstly shuffled the non-overlapping blocks. Then the $(r,v)$ pairs in each block are shuffled to further protect the position information. Finally, the $v$ and $DC$ values are substituted to protect the value information. 

The $\mathcal{K}_p$ is used to encrypt the position information of image. In JPEG-domain, it contains the block permutation, intra-block permutation and the bitstream in one fix length. In detail, a pseudo-random and a stream-cipher generator and several keys are used to protect the position information, i.e., $\mathcal{K}_p$ $= \{\textsf{RandPerm}$, $\textsf{StmCiph}$, $\{ key_{blo*}\}_{* \in \{Y,U,V \}}$, $\{key_{inblo*} \}_{* \in \{Y,U,V \}}$, $\{key_{dc*}\}_{* \in \{Y,U,V \}}\}$. For simplicity, all the following $*$ represent an element in $\{Y, U, V\}$

The secret key $\{key_{blo*}\}$ is used to permute the blocks in an image from the range $[1,\dots,blknum_*]$, the $blknum_*$ is the number of non-overlapping blocks in the corresponding color component. The random permutation is generated as follow:

\begin{equation}
pmtb_{*} \leftarrow \textsf{RandPerm}(key_{blo*}, [1,\dots,blknum_{*}], IID).
\end{equation}

The secret keys $\{key_{inblo*} \}$ are used to generate random permutations to shuffle $(r,v)$ pairs in blocks. The random permutations of the three components are generated as follows:

\begin{equation}
\begin{array}{c}
\{pmtp_{*j}\} \leftarrow \textsf{RandPerm}(key_{inblo*},\\\ [1,\dots, blksize_{j_*}], IID,j_*),

\end{array}
\end{equation}

\noindent where $\{blksize_{j_*}\}$ means the number of $(r,v)$ pairs in $j_*$-th block, $j_* \in [1,\dots,blknum_*]$.

The bit-length information is useful for retrieval. For both protection and utilization, we need to control the bit-length of encrypted $DC$, it is difficult to generate the bitstream to encrypt $DC$ before encryption. In this case, we directly use $key_{dc*}$ to generate the encrypted $DC$, the random bitstream is generated as follow:

\begin{equation}
\begin{array}{c}
\{bitdc_{*j} \} \leftarrow \textsf{StmCiph}(key_{dc*},\\\ [1,\dots, blksize_{j_*}], IID, j_*).

\end{array}
\end{equation}

Accordingly, a pseudo-random permutation generator and several secret keys are used to protect the value information, i.e., $ \mathcal{K}_v=\{\textsf{RandPerm}$, $\{ key_{v*} \}$, $\{key_{l*} \} \}$.

The secret keys $\left\{ key_{v*} \right\}$ are utilized to generate random permutations to substitute the value of $v$ in all the blocks. As the most absolute value of $v$ is less than 10, the random permutations are generated as follows:

\begin{equation}
\left\{ pmtv_{*,\#} \right\} \leftarrow \textsf{RandPerm}(key_{v*}, [-10, -1] \cup [1,10]),
\end{equation}

\noindent here $\# \in \{1,\dots,N_{pmt1}\}$, $N_{pmt1}$ is the number of the random permutations for each color component. The permutations ignore 0 due to the limitation of JPEG-decoding.

The secret keys $\{ key_{l*} \}$ are used to generate the random permutations to substitute the $g_{DC}$. As most $g_{DC}$ is less than 10, the random permutations are generated from the range [0,9] as follows:

\begin{equation}
\{ pmtDCL_{*,\#} \} \leftarrow \textsf{RandPerm}(key_{l*}, [0, 9]),
\end{equation}

\noindent here $\# \in \{1,\dots,N_{pmt2}\}$, $N_{pmt2}$ is the number of the random permutations for each color component. The encryption on $DC$ is determined by both $\{bitdc_{*j} \}$ and $\{ pmtDCL_{*,\#} \}$

It should be noted that the $\mathcal{K}_v$ is unique for each owner, but the $\mathcal{K}_p$ is one-time-pad for each image.

\subsubsection{Image Outsourcing}\label{sec:Imgoursource}

$(C$, $ImgPosKey$, $ImgValKey)$ $\leftarrow$ $ImgEnc($$I$, $IID$, $\mathcal{K}_v$). As presented above, three steps are contained in the image encryption including block permutation, intra-block permutation, and value substitution. For each step, we present a sub-algorithm to specify its process(see Algorithm \ref{alg: BloPerm}, \ref{alg: IntBloPerm} and \ref{alg: valuesubsti}).

\begin{algorithm}[t]
	\caption{BlockPermut}
	\label{alg: BloPerm}
	\begin{algorithmic}[1]
		\REQUIRE
		Image $I$, the corresponding $IID$ and secret keys $\left\{ key_{blo*} \right\}$
		\ENSURE
		Encrypted image $I'$, $\{pmtb_*\}$ \\
		
		\STATE Parse the image, and denote the total number of blocks in image $I$ as $blknum_{*}$
		
		\STATE Generate the secret permutation $pmtb_*$ whose size is $blknum_{*}$
		
		\STATE Denote the blocks in $I$ as $blk$, denote the blocks in $I'$ as $blk'$
		
		\FOR{$\forall * \in Y,U,V$}
		\FOR{$i = 1:blknum_{*}$}
		\STATE $blk'_* \leftarrow blk_*[pmtb_*[i]]$
		\ENDFOR
		\ENDFOR
		
	\end{algorithmic}
\end{algorithm}
\begin{algorithm}[t]
	\caption{IntraBlockPermut}
	\label{alg: IntBloPerm}
	\begin{algorithmic}[1]
		\REQUIRE
		Image $I$, the corresponding $IID$ and secret keys $\left\{ key_{blo*} \right\}$
		\ENSURE
		Encrypted image $I'$,$\left\{pmtp_* \right\}$\\
		
		\STATE Parse the image, and get the blocks denoted by $blk_*$
		
		\FOR{$* \in \left\{Y,U,V \right\} $}
		
		\FOR {$blk_{*j} \in blk_*$}
		\STATE Generate the secret permutation for j-th block $blk_{*j}$ size of $blksize_{j*}$ as $pmtp_{*j}$
		\FOR{ $blk_{*j}[i] \in blk_{*j}$}
		\STATE $blk'_{*j}[i] \leftarrow blk_{*j}[pmtp_{*j}[i]]$
		\ENDFOR
		\ENDFOR
		\STATE Denote all the $pmtp_{*j}$ as $\left\{pmtp_* \right\}$
		\ENDFOR
		
	\end{algorithmic}
\end{algorithm}

\begin{algorithm}[t]
	\caption{ValueSubstitution}
	\label{alg: valuesubsti}
	\begin{algorithmic}[1]
		\REQUIRE
		Image $I$ and secret keys $\{ key_{v*} \}$, $\{ key_{l*}\}$ and $key_{dc*}$
		\ENSURE
		Encrypted image $I'$, $\{pmtv_{*,\#}\}$ and $\{pmtDCL_{*,\#}\}$, $\{bitkey_*\}$
		
		\STATE Generate the secret permutations $pmtv_{Y,\#}$, $pmtv_{U,\#}$, $pmtv_{V,\#}$, where $\# \in \left\{ 1,\dots, N_{pmt1}\right\}$; Each permutation table is 20-dim, which is a random permutation of $[-10,-1]$ $\cup$ $[1,10]$.
		
		\STATE Generate the secret permutations $pmtDCL_{Y,\#}$, $pmtDCL_{U,\#}$, $pmtDCL_{V,\#}$, where $\# \in \{1,\dots,N_{pmt2} \}$. Each permutation table is 10-dim, which is a random permutation of $[0,9]$.
		
		\STATE Generate six sequences $sqnt_{1*}$ and $sqnt_{2*}$. The length of sequences are equal to the block amount of the image $I$ and the element of $sqnt_{1*}$ are the repeat of $E_{N_{pmt1}}$. For instance, if the $N_{pmt1} = 5$, and the image have 12 blocks, the $sqnt_{1*} = \{1,2,3,4,5,1,2,3,4,5,1,2 \}$. Similarly, $sqnt_{2*}$ are generated by $N_{pmt2}$.
		
		\STATE Parse the image and get the $\{\{(r_{ij}^*,v_{ij}^*)_{j=1}^{\{*blksize_{i*}\}_{i=1}^{blknum_*}}\}$, $\{ DC_i^* \}_{i=1}^{blknum_{*}}$.
		
		\FOR{$\forall * \in Y,U,V$}
		\FOR{$i = 1:blknum_{*}$}
		
		\FOR{$j = 1:blksize_{i*}$}
		\STATE $encv_{ij}^* = pmtv_{*,sqnt_{1*}[i]}[v_{ij}^*]$
		\ENDFOR
		
		\STATE Generate a random bitstream $bitdc_{*i}$, then only save last $pmtDCL_{*,sqnt_{2*}}[g_{DC_i^{*}}]$ bit as the $encDC_i^*$.
		
		\STATE Compute $bitkey_{i}^{*}$ $=$ $encDC_i^*$ $\oplus$ $DC_i^*$
		
		
		
		\ENDFOR
		\STATE Denote all the $bitkey_i^*$ as $\{bitkey_*\}$
		\STATE The  $\{\{(r_{ij}^*,encv_{ij}^*)_{j=1}^{\{*blksize_{i*}\}_{i=1}^{blknum_*}}\}$ and $\{encDC_i^* \}_{i=1}^{blknum_{*}}$ compose encrypted image $I'$.
		\ENDFOR
		
	\end{algorithmic}
\end{algorithm}
\begin{algorithm}[t]
	\caption{ImgEnc}
	\label{alg: OwnImgEnc}
	\begin{algorithmic}[1]
		\REQUIRE
		Image $I$, the corresponding $IID$ and $K_v$
		\ENSURE
		Encrypted image $C$, $ImgPosKey_{(OID,IID)}$ and $ImgValKey_{OID}$\\
		
		\STATE Randomly generate $\mathcal{K}_p$
		
		\STATE $(I'_i,pmtb_*) = \textsf{BlockPermut}(I_i,\left\{key_{blo*}\right\})$
		
		\STATE $(I''_i,\left\{pmtp_* \right\}) = \textsf{IntraBlockPermut}(I'_i, key_{inblo*,\#})$
		
		\STATE $(C,\{bitkey_*\},\{pmtDCL_{*,\#}\},\{pmtv_{*,\#}\}) = $ \\$\textsf{ValueSubstitution}(I''_i,key_{v*}, key_{l*}, key_{dc*})$
		
		\STATE Denote the ($pmtb_*$, $\left\{pmtp_* \right\}$, $\{bitkey^*\}$) as the $ImgPosKey_{(OID,IID)}$, denote the ($\{pmtDCL_{*,\#}\}$, $\{pmtv_{*,\#}\}$) as the $ImgValKey_{OID}$
	\end{algorithmic}
\end{algorithm}

As presented by algorithm \ref{alg: BloPerm} and \ref{alg: IntBloPerm}, we generate random permutations to shuffle the block position and intra-block $(r,v)$ pairs. Notably, the permutations on plaintext image is equal to execute the $\textsf{EncPerm}$ algorithm where one of the input can seem as $E$. The Algorithm \ref{alg: valuesubsti} encrypts the $v$ and $DC$ by substituting values with multiple tables. In this way, the same value at different positions can be substituted with different values, which helps to resist the statistic attacked \cite{chen2004symmetric,chai2017novel}. What's more, it helps to resist the collusion between image owner and CS as shown in section \ref{sec:section6}.

As shown in Algorithm \ref{alg: OwnImgEnc}, we denote the ($pmtb_*$, $\left\{pmtp_* \right\}$, $\{bitkey_*\}$) as the $ImgPosKey_{(OID,IID)}$. For simplicity, the $OID$ and $IID$ will be omitted when there is no ambiguity or in general reference in the rest of the paper. Please note that all the encryption methods are high-efficiency bit computation or vector operation. After encryption, the owner should send the encrypted image $C$ to the CS, and send the corresponding encryption key $ImgPosKey_{(OID, IID)}$ to the KMC.

\subsubsection{User authorization}\label{sec:authorization}

When one image owner with identity $OID$ wants to authorize the user, he will give the user $ImgValKey_{OID}$ for the retrieval and decryption. The authorization information will also be known by CS and KMC. The further operations during authorization will be introduced in subsection \ref{subsec:ImgPosKeyPro}.

\subsection{Authorized user Side}
The authorized user wants to search similar images from the owners who authorize him. As shown in algorithm \ref{alg: TrapGen}, the user just needs to encrypt the query image with $ImgValKey$ he gets from the owners. It is noteworthy that the encrypted query is protected by the \textsf{BlockPermut} at last, which means the relationship of blocks is destroyed. After encryption, the authorized user only needs to send all the encrypted queries with corresponding $OID$ as the trapdoor to CS.

\begin{algorithm}[t]
	\caption{TrapGen}
	\label{alg: TrapGen}
	\begin{algorithmic}[1]
		\REQUIRE
		Image $I$, $\mathcal{K}_v$
		\ENSURE
		Encrypted query images $\{C_{OID}\}$\\
		
		\FOR{each $OID$ who authoring}
		\STATE Randomly generate a $\mathcal{K}_p$ and a $\left\{key'_{blo*}\right\}$
		
		\STATE $(I'_i,\sim) = \textsf{BlockPermut}(I_i,\left\{key_{blo*}\right\})$
		
		\STATE $(I''_i,\sim) = \textsf{IntraBlockPermut}(I'_i, key_{inblo*})$
		
		\STATE $(I'''_i,\sim) = \textsf{ValueSubstitution}(I''_i,key_{v*}, key_{l*}, key_{dc*})$
		
		\STATE $(C_{OID},\sim,\sim,\sim) = \textsf{BlockPermut}(I'''_i, \left\{key'_{blo*}\right\})$
		
		\ENDFOR
	\end{algorithmic}
\end{algorithm}

\subsection{Cloud Side}\label{subsec:FeaExtra}

After the owners upload the encrypted images, for the efficiency of retrieval, the CS will extract high-quality encrypted feature and further build the index for images in the database. As the process of index building is same to that in the plaintext situation, we here focus on encrypted feature extraction and aggregation process.

\subsubsection{Global feature extraction from encrypted DC}\label{sec:globalfeature}

The $g_{DC}$ is extracted to represent $DC$ information of encrypted image \cite{schaefer2001jpeg}. As most values of $g_{DC}$ are concentrated in [0,9], the CS can extract a 10-dim feature, in which the $j$th-dim represents the number of $g_{DC}$ whose value equals to $j$. The Y,U,V further form a 30-dim feature vector $f^{DC}$.

\subsubsection{Aggregation feature extraction from the encrypted AC}\label{sec:localfeature}

Different from $DC$, it is difficult to represent the AC values in a block effectively by a single number and it makes the feature aggregation an indispensable step. Inspired by \cite{xia2019boew}, we use the typical BOW model to aggregate the features. The kernel observation here is that the encrypted histogram can still be used to compute the distance and $k$-means method BOW \cite{sivic2003video} uses is robust to the element permutation. Accordingly, the aggregation for encrypted AC values consists of the following three steps:

\emph{(i) Local histogram extraction}. A 40-dim vector is extracted to represent the feature of encrypted AC in each block, which is composed of three parts as formula \ref{eq:ACFeature}.

\begin{equation}
\label{eq:ACFeature}
f_{ACLocal} = Hist_{s} \parallel Hist_{v} \parallel Hist_{r}.
\end{equation}

The $Hist_{s}$ contains the static information of $(r,v)$ pairs, including the number of $(r,v)$ pairs, the mean and standard deviation of $r$. The $Hist_{v}$ is the distribution information of $v$. In detail, the value of $21$-dim is the number of occurrences of $v$ values in the block range from $[-10,10]$, and the other $2$-dim represents the $v$ values more than $10$ or less than $-10$. The $Hist_{r}$ is the value information of $r$. The $14$ biggest value of $r$ form the vector in descending order. If the number of $(r,v)$ pairs is less than $14$, the unfilled elements of $Hist_{r}$ will be filled by $-1$.

\emph{(ii) Vocabulary generation}. Cluster all the local features into $k$ classes with the $k$-means clustering algorithm. The $k$ cluster centers are defined to be the encrypted visual words that make up the vocabulary. It should be noticed the features extracted from Y, U, and V are clustered independently as they have different properties naturally. The selection of $k$ is always a difficult problem, however, the methods like x-means \cite{pelleg2000x} or gap statics \cite{tibshirani2001estimating} can effectively cope with the problem. What's more, we will show the retrieval accuracy of our scheme is quite robust to $k$ in Fig. \ref{fig:DiffK}.

\emph{(iii) Histogram calculation}. After generating the vocabulary, all the local histograms in an image are represented by their nearest visual words. As a result, each image is represented by a feature vector $f = (f_i)_{i=1}^k$. A "scaled tf-idf" \cite{chum2008near} trick is further implemented to optimize the feature $f^{Y}$, $f^{U}$ and $f^{V}$.

Finally, the image identities and the feature vector make up a linear index. It is easy to see that the feature vectors are encrypted but the common index building schemes (e.g., tree index \cite{muja2014scalable}) can be further used.

\subsubsection{Search operation}

When the CS gets the trapdoor generated by the authorized user, it will extract the same format feature as that from images in the dataset. If the query is limited in a single owner, the CS will calculate the feature with the corresponding visual words. Detailedly, the distance are calculated as formula \ref{eq:discompute}, where $D(\cdot,\cdot)$ means manhattan distance.

\begin{equation}
\label{eq:discompute}
\begin{array}{c}
Dis(I_1,I_2) = \alpha_1 D(f_{I_1}^{DC},f_{I_2}^{DC}) + \alpha_2 D(f_{I_1}^{Y},f_{I_2}^{Y})\\
+ \alpha_3 D(f_{I_1}^{U},f_{I_2}^{U}) + \alpha_4 D(f_{I_1}^{V},f_{I_2}^{V})
\end{array}
\end{equation}

Follow the experience and experiments, we set $\alpha_1 = 0.1$, $\alpha_2 = 0.5$, $\alpha_3 = \alpha_4 = 0.2$.

\subsubsection{Search operation from multiple image owners}\label{sec:IdxForMulSource}
The encrypted feature extraction method described above is still valid during the retrieval in that the manhattan distance will not change if we execute the same permutation on the elements of feature vectors. When the multiple permutation tables are used, the high-frequency values will be randomly substituted to $N_{pmt}$ different value, where $N_{pmt}$ is the number of tables. It means if the value frequency distribution of two images is similar, the frequency of encrypted images will still have an extent of similarity, although the frequency becomes smoother with the increment of $N_{pmt}$. The difference of distances becomes smaller, however, the size relationship is still basically kept which is demonstrated experimentally.

It further implies that if the images are encrypted by the permutation tables which have the same $N_{pmt}$, the distances after encryption are still at the same level even though different permutations are utilized for encryption. It means the formula \ref{eq:encencdis} set on if we use the above encryption methods, where $I^{enc1}$ and $I^{enc2}$ means the image encrypted with different $ImgValKey$ and $ImgPosKey$.

\begin{equation}
\label{eq:encencdis}
\begin{array}{c}
Dis(I_{1}^{enc1},I_{2}^{enc1}) \approx Dis(I_{1}^{enc2},I_{2}^{enc2})
\end{array}
\end{equation}

Please note that the indispensable aggregation schemes will also infect the distance relationship. To keep the distance can be directly compared, the same cluster number $k_g$ are used to cluster the images from each image owner. The choice of $k_g$ will be discussed in subsection \ref{subsec:multi-source}.

When the query contains multi-sources, the CS will calculate the feature based on each global visual word, the distance got from different sources will be directly compared together and images with smaller distance will be returned.

\subsection{KMC side}

After CS gets similar images, it sends the ($\{IID\}$, $UID$, $OID$) to KMC. Here, we follow the operation in \cite{ferreira2017practical}, the CS sends encrypted images $\{C\}$ to the querier and KMC sends the corresponding $\{ImgPosKey\}$. The user will decrypt the retrieval results according to the key he has got. Notably, it makes users have to interact with the KMC and actually leads to two rounds of interaction. We will make up the drawback in the next section.

\section{Advanced Scheme}\label{sec:section5}

In the previous section, we give the scheme that can support secure retrieval from multi-source. However, as the $ImgPosKey$ is directly sent to KMC, it will be fragile to face the conspiracy between CS and KMC. Based on the same reason, the CS and KMC have to interact with the user respectively, which leads to extra interaction for the user. In this section, we firstly propose the scheme for protecting $ImgPosKey$ to enhance the security and reduce interaction rounds, then the strategy for group union scene is further given.

\subsection{key protection}

Inspired by formula \ref{eq:decencenc}, we design a safer scheme with little computation increment during authorization. Briefly speaking, to hidden $ImgPosKey$, each owner with identity $OID$ constructs a series of key called as $UserKey_{OID}$. Like formula \ref{eq:decencenc}, $ImgPosKey$ plays the role of $K_2$, $UserKey_{OID}$ plays the role of $K_1$. During authorization, the owner will generate and send a random $UserKey_{UID_{OID}}$ which is the same format with $UserKey_{OID}$ to the user. For simplicity, the $OID$ will be omitted in $UserKey_{UID_{OID}}$.

Briefly speaking, the $UserKey_{UID}$ plays the role of $K$ in formula \ref{eq:decencenc}. The work of computing $\textsf{DecPerm}(K_2, K_1)$ will be undertaken by the owner during image outsourcing, and work of computing $\textsf{EncPerm}(K_1, K)$ will be undertaken by the owner during authorization, and the result will be stored by KMC. KMC undertakes the computation of formula \ref{eq:decencenc} during the query. Here we give the construction method of $UserKey_{OID}$.

\subsubsection{UserKey generation}

To encrypt the $ImgPosKey$, owner has to consider all the situation. In detail, the owner needs to generate the key for encrypting the block-permutation key, intra-block permutation key and the stream cipher. Accordingly, a pseudo-random permutation generator, a stream-cipher generator and several secret keys are included in the $UserKey$, i.e., $K_u$ = $\{$ $\textsf{RandPerm}$, $\textsf{StmCiph}$, $\{key_{Ublo*}$ $\}$, $\{key_{Uinblo*}$ $\}$, $\{key_{Udc*}\}$ $\} $

Here, the secret keys $\{ key_{Ublo*} \}$ are utilized to generate random permutations that are used to encrypt the inter-block permutation keys. As the length of inter-block permutation is determined by the size of images, it is difficult to consider all the situations. However, it can be remedied with a series of permutations whose length is the exponential of two. For simplicity, we here assume the length of images denoted as $\{CommSize\}$ are all under consideration, and the random permutations are generated as follows:

\unskip
\begin{equation}
\{ Upmtb_{*\#} \} \leftarrow \textsf{RandPerm}(key_{Ublo*}, \{CommSize\}),
\end{equation}

\noindent where $\# \in \{1,\dots,|\{CommSize\}|\}$, here $|\{CommSize\}|$ is the cardinality of set $\{CommSize\}$. Similarity, the secret keys $\left\{ key_{Uinblo*} \right\}$ are used to protect the intra-block permutations. As the amount of $(r,v)$ pairs is in [1, 63], it can be generated as follows:

\unskip
\begin{equation}
\{ Upmtp_{*\#} \} \leftarrow \textsf{RandPerm}(key_{Uinblo*}, [1,\dots,63]),
\end{equation}

\noindent where $\# \in \{1,\dots,63\}$. At last, the $key_{Udc*}$ is used to encrypt the bitstream which is computed to decrypt $DC$. The stream-ciphers are generated as follows:

\unskip
\begin{equation}
\{ UbitKey_{*\#} \} \leftarrow \textsf{StmCiph}(key_{Udc*}, [1,\dots,10]),
\end{equation}

\noindent where $\# \in \{1,\dots,10\}$. The whole ($\{ Upmtb_{*\#} \}$, $\{ Upmtp_{*\#} \}$ and $\{ UbitKey_{*\#} \}$) is denoted as the $UserKey$.

\subsubsection{Encryption on image position key}\label{subsec:ImgPosKeyPro}

Different from subsection \ref{sec:Imgoursource}, after the encryption of an image, the image owner won't directly send $ImgPosKey$ to KMC. As shown in Algorithm \ref{alg: ImgKeyEnc}, the image owner will use the $UserKey_{OID}$ to encrypt the $ImgPosKey$, and then send $ImgPosKey'$ $=$ $\textsf{ImgKeyEnc}(ImgPosKey, UserKey_{OID})$ to KMC. Notably, $UserKey$ is unique to each image owner, and it will not be exposed to anyone else. The encrypted key $ImgPosKey'$ are the same format with $ImgPosKey$.

\begin{algorithm}[t]
	\caption{ImgKeyEnc}
	\label{alg: ImgKeyEnc}
	\begin{algorithmic}[1]
		\REQUIRE
		$ImgPosKey_{IID}$, $UserKey_{OID}$
		\ENSURE
		$ImgPosKey'_{IID}$
		\FOR{$\forall$ $pmtp_{*j} \in pmtp_{*}$}
		\STATE seek the same length permutation $Upmtp_{*j}$ in $Upmtp_{*\#}$
		\STATE $pmtp'_{*j}$ = $\textsf{DecPerm}(pmtp_{*j}, Upmtp_{*j})$
		\ENDFOR
		
		\FOR{$\forall$ $* \in \left\{H,S,V \right\}$}
		\STATE seek the same length permutation $Upmtb_{*j}$ in $Upmtb_{*\#}$
		\STATE $pmtb'_{*}$ = $\textsf{DecPerm}(pmtb_{*}, Upmtb_{*j})$
		\ENDFOR
		
		\FOR{$\forall$ $bitkey_{i*} \in \{bitkey_*\} $}
		\STATE seek the same length bit-stream $Ubitkey$ in $Ubitkey'_{*\#}$
		\STATE $bitkey'_{i*}$ = $bitkey_{i*} \oplus Ubitkey$
		\ENDFOR
		
		\STATE Denote the ($pmtb'_{*}$, $\{pmtp'_{*}\}$, $\{bitkey'_{i}\}$) as the $ImgPosKey'$.
		
	\end{algorithmic}
\end{algorithm}
When the owner authorizes an user, he randomly generates a $UserKey_{UID}$ and send to the authorized user. Then owner will use $UserKey_{UID}$ to encrypt the $UserKey_{OID}$ for following retrieval. As Algorithm \ref{alg: UserKeyEnc} shows, the owner computes $\textsf{UserKeyEnc}(UserKey_{OID},UserKey_{UID})$, then sends the result $UserKey_{(OID,UID)}$ to KMC.

\begin{algorithm}[t]
	\caption{UserKeyEnc/UserKeyDec}
	\label{alg: UserKeyEnc}
	\begin{algorithmic}[1]
		\REQUIRE
		$UserKey_{OID}$, $UserKey_{UID}$
		\ENSURE
		$UserKey_{(OID, UID)}$
		
		\FOR{$\forall$ $Upmtb_{*j}^{OID}$ $\in$ $Upmtb_{*\#}^{OID} $}
		\STATE seek the same length permutation $Upmtb_{*j}^{UID}$ in $Upmtb_{*\#}^{UID}$
		\STATE $Upmtb_{*j}$=$\textsf{EncPerm/DecPerm}(Upmtb_{*j}^{OID},Upmtb_{*j}^{UID})$
		\ENDFOR
		
		\FOR{$\forall$ $Upmtp_{*j}^{OID}$ $\in$ $Upmtp_{*\#}^{OID} $}
		\STATE seek the same length permutation $Upmtp_{*j}^{UID}$ in $Upmtp_{*\#}^{UID}$
		\STATE $Upmtp_{*j}$=$\textsf{EncPerm/DecPerm}(Upmtp_{*j}^{OID},Upmtp_{*j}^{UID})$
		\ENDFOR
		
		\FOR{$\forall$ $Ubitkey_{*j}^{OID}$ $\in$ $Ubitkey_{*\#}^{OID} $}
		\STATE seek the same length bitstream $Ubitkey_{*j}^{UID}$ in $Ubitkey_{*\#}^{UID}$
		\STATE $Ubitkey_{i}$=$Ubitkey_i^{OID} \oplus Ubitkey_{i}^{UID}$
		\ENDFOR
		
		\STATE Denote ($\{Upmtb_{*\#} \}$, $\{ Upmtp_{*\#}\}$, $\{UbitKey_{*\#}\}$) as $UserKey_{(OID, UID)}$
		
	\end{algorithmic}
\end{algorithm}

When CS asks the secret key from KMC, KMC will compute $EncImgPosKey'_{IID} =  \textsf{ImgKeyEnc}(ImgPosKey'_{IID}$, $UserKey_{(OID,UID)})$, and send back to CS. CS will finally send encrypted images $\{C\}$ and $EncImgPosKey'_{IID}$ to the authorized user. From formula \ref{eq:decencenc}, it is easy to note that the authorized user can get the encryption key with the help of $ImgValKey$, $EncImgPosKey'$, and $UserKey_{UID_{OID}}$.

\subsection{Group Union}\label{subsec:GroupUnion}
\begin{figure}[tb]
	\centering
	\includegraphics[width=1.0\linewidth]{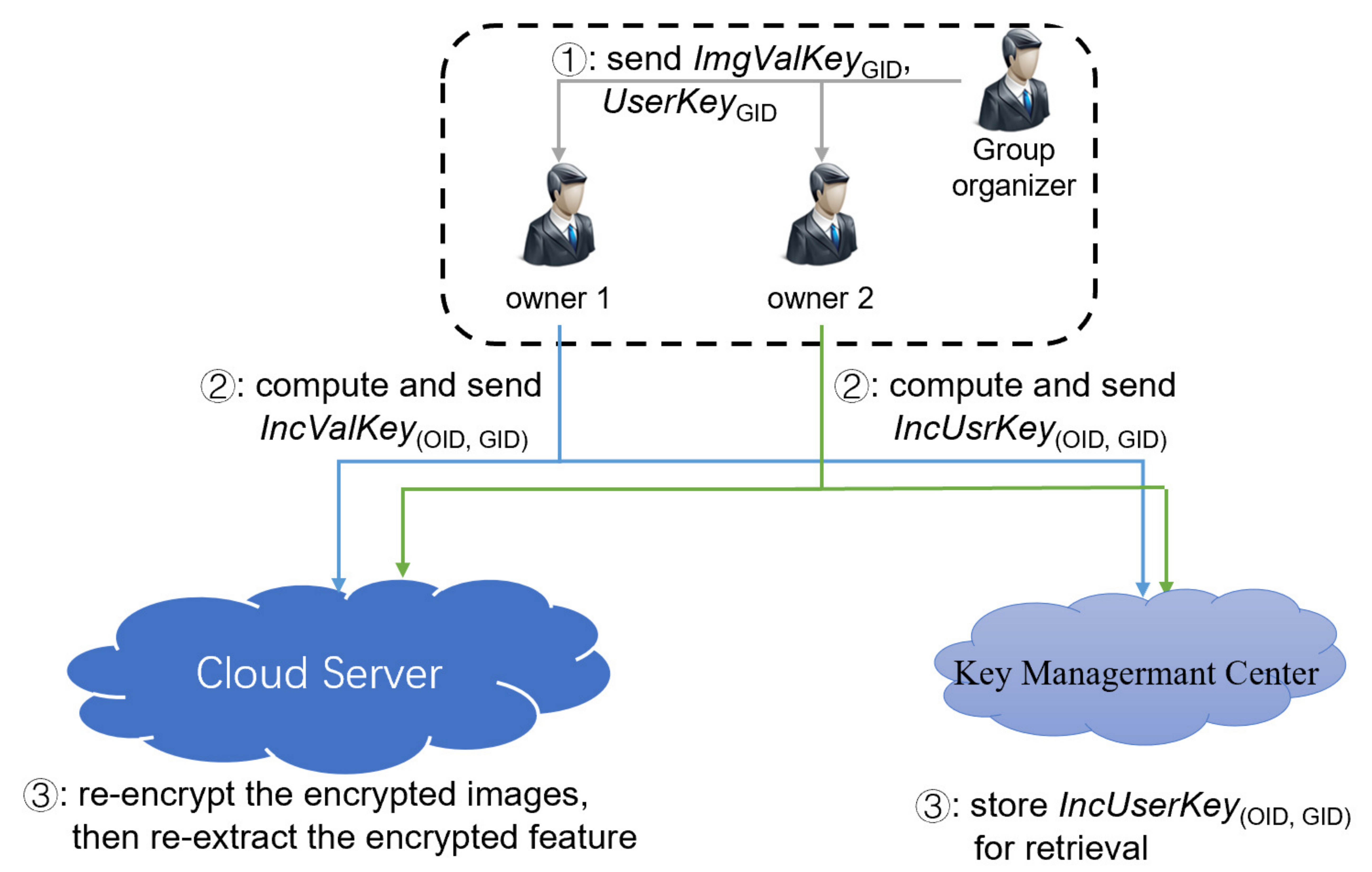}
	\caption{The process of group union}
	\label{fig:GroupUnion}
\end{figure}

Inspired by \cite{ferreira2017practical}, the situation that owners unite as a group is further considered. In \cite{ferreira2017practical}, the creator creates a repository in the CS, and the member join in should use the $ImgValKey$ creator set to encrypt their images, and further upload them into the repository. It will lead to plenty of extra consumption when the owner wants to join in different groups. Inspired by formula \ref{eq:encenc}, we accomplish the union by some increment keys. The process of union is shown as figure \ref{fig:GroupUnion}.

When the owners unite as the group, they should firstly choose  a trusty member or third party as the group organizer. Similar to the image owner, the group organizer will randomly generate $UserKey_{GID}$ and $ImgValKey_{GID}$, then send them to all the members in the group. After getting these information, to meet the demand of decryption for users who authorized by the group organizer, the members will compute $IncUsrKey_{(OID,GID)}$ = $\texttt{UserKeyDec}(UserKey_{OID}$, $UserKey_{GID})$, and send the result to KMC. Similarly, in order to meet the demand of retrieval, the members will compute all the $\textsf{DecPerm}($$pmtv_{*,\#}^{OID}$, $pmtv_{*,\#}^{GID})$ and $\textsf{DecPerm}($ $pmtDCL_{*,\#}^{OID}$, $pmtDCL_{*,\#}^{GID})$, the results can be denoted as $IncValKey_{(OID,GID)}$, and image owner will send it to CS. Based on formula \ref{eq:encenc}, the CS could further execute the encryption on encrypted images, then the CS further executes the same operation in subsection \ref{subsec:FeaExtra}. The $(r,v)$ part linear index of images for one image owner in CS will be finally built like Table \ref{tab:LinearIndex}.

The group organizer undertakes the task of authorization. Besides the $UserKey_{UID_{GID}}$, the group organizer should use the symmetric encryption to avoid the potential collusion risk (shown in subsection \ref{subsec:colludeCSandMember}). Here we briefly use the AES (Advanced Encryption Standard) \cite{daemen2001reijndael}. In this case, after the authorization, the group organizer will send the $UserKey_{(UID_{GID}, GID)}$ and $\mathcal{K}_k^{(GID,UID)}$ to KMC.

During the retrieval from the group, CS will compute the similarity with the features belong to the group and then send ($GID$, $UID$, $\{OID\}$, $\{IID\}$) to KMC. Based on $GID$ and $OID$, KMC can seek the corresponding $IncUsrKey$. Then, based on $IID$, KMC will firstly compute $\textsf{ImgKeyEnc}($ $ImgPosKey,$ $IncUsrKey)$. Then, same to single owner, KMC can compute and get $EncImgPosKey'$ with the help of $UserKey_{(GID,UID)}$. At last, AES encryption based on $\mathcal{K}_k^{(GID,UID)}$ will be executed on $EncImgPosKey'$, and the results will be sent back to CS. It is easy to notice that the user can finish the decryption with the owned keys.

\begin{table*}[tb]
	\setlength{\abovecaptionskip}{0pt}
	\setlength{\belowcaptionskip}{10pt}
	\centering
	\caption{Partial linear index built for one image owner}
	\label{tab:LinearIndex}
	\begin{tabular}{|l|c|c|c|c|c|}
		\hline
		\multirow{2}{*}{\begin{tabular}[c]{@{}l@{}}Image\\ Identity\end{tabular}} & \multicolumn{5}{c|}{Feature vector aggregated from $(r,v)$ pairs for different authorized users} \\ \cline{2-6} 
		& $Owner_{OID}$               & $Global_{OID}$              & $Group_{GID_{1}}$              & $Global_{GID_{1}}$             & \dots \\ \hline 
		$IID(C_1)$ & $f_{1}^{OID} = \{f_{1j}^{OID}\}_{j=1}^{k_{OID}}$ & $f_{1G}^{OID} = \{f_{1gj}^{OID}\}_{j=1}^{k_g}$ & $f_{1}^{GID_{1}} = \{f_{1j}^{GID_{1}}\}_{j=1}^{k_{GID_{1}}}$ & $f_{1G}^{GID_{1}} = \{f_{1gj}^{GID_{1}}\}_{j=1}^{k_g}$ & \dots        \\ \hline
		\dots     & ...                    & \dots                    & \dots                    & \dots                    & \dots        \\ \hline
		$IID(C_i)$ & $f_{i}^{OID} = \{f_{ij}^{OID}\}_{j=1}^{k_{OID}}$ & $f_{iG}^{OID} = \{f_{igj}^{OID}\}_{j=1}^{k_g}$ & $f_{i}^{GID_{1}} = \{f_{ij}^{GID_{1}}\}_{j=1}^{k_{GID_{1}}}$ & $f_{iG}^{GID_{1}} = \{f_{igj}^{GID_{1}}\}_{j=1}^{k_g}$          & \dots \\ \hline
		\dots     & \dots                    & \dots                    & \dots                    & \dots                    & \dots       \\ \hline
		$IID(C_n)$ & $f_{n}^{OID} = \{f_{nj}^{OID}\}_{j=1}^{k_{OID}}$ & $f_{nG}^{OID} = \{f_{ngj}^{OID}\}_{j=1}^{k_g}$ & $f_{n}^{GID_{1}} = \{f_{nj}^{GID_{1}}\}_{j=1}^{k_{GID_{1}}}$ & $f_{nG}^{GID_{1}} = \{f_{ngj}^{GID_{1}}\}_{j=1}^{k_g}$  & \dots        \\ \hline
	\end{tabular} 
\end{table*}

\subsection{Update operation}

After introducing all the entities, the update operations in JES-MSIR are given here. In detail, We will show the update on images and owners.

\emph{Image addition}: As the section \ref{sec:section3}, based on his $ImgValKey$ and random $ImgPosKey$, the owner can get the encrypted image and sends it to CS. Similarly, based on the $UserKey$, the owner will compute $\textsf{ImgKeyEnc}(ImgPosKey, UserKey)$ and send it to KMC. The CS will execute the feature extraction by existing visual words, and add this image into the index. The KMC will store the key for the following retrieval.

\emph{Image deletion}: The owner with identity $GID$ sends $IID$ to CS and KMC. Then CS should delete the corresponding encrypted image and all the feature in the index, KMC should delete the corresponding $ImgPosKey'_{(OID, IID)}$

\emph{Join group}: The group organizer in the group with identity $GID$ sends its $UserKey_{GID}$ and $ImgValKey_{GID}$ to the new member, then the member, CS, and KMC will execute the same operation in subsection \ref{subsec:GroupUnion}.

\emph{Leave group}: The group organizer in the group with identity $GID$ sends the $OID$ to CS and KMC. Then CS should delete the corresponding image features which are extracted for the owner and group, KMC should delete the $IncUsrKey_{(OID, GID)}$ 

\section{Security Analysis}\label{sec:section6}

Besides the security problems in the PPCBIR \cite{ferreira2017practical,xia2019boew}, the introduction of MSPPIR \cite{zhang2017pic} also brings the conspiracy risk from different entities. The security analysis in the non-collusion assumption, including Ciphertext-Only Attack (COA) and Known-Background Attack (KBA), will be firstly given in subsection \ref{sec:non-collusion}, then we analyze the potential collusion problems in subsection \ref{sec:collusion}, finally the security comparison with previous schemes in MSPPIR are given in subsection \ref{subsec:securitycompare}.

\subsection{Security with no Collusion}\label{sec:non-collusion}
\subsubsection{Security under COA model}

In the COA model, the adversary can only get the ciphertext. As the images are all stored in the CS, we here mainly consider the potential leakage in CS side. Follow the universally composition framework, for formal statements, the functionality $\mathcal{F}$ and the corresponding information leakages of our scheme under the COA model are summarized in Fig. \ref{fig:FunctionLeakage}. The interaction between CS and other entities during executing our scheme is defined as the real experiment. In this case, the honest-but-curious CS is the potential adversary $\mathcal{A}$. In the ideal experiment, the simulator $\mathcal{S}$ is defined as the one that can simulate the view of $\mathcal{A}$ by using functionality $\mathcal{F}$. The proposed scheme is proved secure once the two experiments are indistinguishable. Here, for simplicity, we mainly focus on the security analysis of image content of its feature.

\begin{figure*}[ht]
	\centering
	\fbox{%
		\parbox{0.95\linewidth}{%
			\begin{small}
				
				$\textsf{}$ The mainly ideal functionality $\mathcal{F}$ of our scheme as well as the corresponding information leakages.
				\medskip
				\begin{enumerate}[(i)]
					
					\item $\mathcal{F}.\textsf{StoreImage}(\mathcal{I}, UID, \mathcal{IID}, \mathcal{K}_p, \mathcal{K}_v)$:
					\begin{itemize}[]
						\item \textbf{Functionality}. Each image owner encrypts all his images in $\mathcal{I}$, and generates a set of encrypted images $\mathcal{C}$. Next, each image owner uploads $\mathcal{C}, UID, \mathcal{IID}$ to the CS.
						\item \textbf{Storage leakage}. The information leaked here includes $\mathcal{C}, \mathcal{IID}, UID$ and the size of each images, and the total number of images. What's more, the CS know the corresponding blocks are encrypted by the same valuesubstitution table.
					\end{itemize}
					\medskip
					
					\item $\mathcal{F}.\textsf{Union}(\mathcal{OID},GID,\{IncValKey\})$:
					\begin{itemize}[]
						\item \textbf{Functionality}. Image owners union as a group, and sends the $IncValKey$ to the CS.
						\item \textbf{Relation leakage}. The information leaked here includes the $OID$ in the same group, and the increment key itself. What's more, the CS can compute the difference of $ImgValKey$ belong to the owner in the same group. 
					\end{itemize}
					
					\item $\mathcal{F}.\textsf{IndexGen}(\mathcal{C}, \mathcal{UID}, \mathcal{GID})$:
					\begin{itemize}[]
						\item \textbf{Functionality}. CS extracts local histograms from images blocks belong to each image owner, and constructs the vocabulary by cluster algorithm, and calculates the feature vectors for each image in $\mathcal{C}$ based on the corresponding $UID$ and $GID$ like Table \ref{tab:LinearIndex}.
						\item \textbf{Feature leakage}. The information leaked here includes the encrypted local histograms, the similarities and distributions of local histograms belong to the same source.
					\end{itemize}
					\medskip
					
					\item $\mathcal{F}.\textsf{Query}(\{I_q, \mathcal{UID}, \mathcal{GID}\})$:
					\begin{itemize}[]
						\item \textbf{Functionality}. Authorized user encrypts the query image, and submits the encrypted images to cloud server as trapdoor. The CS execute the similarity calculation, and get the $\{IID\}$ of similar image, and ask KMC the corresponding decryption key. The CS finally return all the $\{UID/GID$, $C$, $EncImgPosKey'\}$ to querier.
						\item \textbf{Query leakage}. The information leaked here includes the encrypted query images and the similarity between the images in the database. The information that encrypted images are encrypted from the same image is also leaked.
					\end{itemize}
					\medskip
					
				\end{enumerate}
			\end{small}
		}
	}
	\caption{The functionality $\mathcal{F}$ and the information leakage in our framework }
	\label{fig:FunctionLeakage}
\end{figure*}

\textbf{Theorem 1.} \emph{Our scheme is secure against an honest-but-curious probabilistic polynomial time (PPT) adversary under the COA model. The security strength depends on the image size, and the number of permutations in ImgValKey.}

\textbf{Proof.}

\begin{itemize}[]
	\item \emph{Security of image content}. As shown in Fig. \ref{fig:FunctionLeakage}, the simulator $S$ simulates a set of images $\mathcal{I^{\mathcal{S}}}$, and the corresponding identity set $\mathcal{IID^{\mathcal{S}}}$ according to the storage leakage. The total number of images and the size of each image are inevitable leaked. However, $\mathcal{S}$ can only fill the images with randomly generated pixels. As described above, JPEG-format images are mainly consisted of the $DC$ values and $(r,v)$ pairs in $Y,U,V$ components. As stated in subsection \ref{sec:Imgoursource}, the above information is protected respectively by the substitutions and random permutations with different keys. The $v$ information between the real images and simulated ones are indistinguishable according to the property of random permutation. For a random permutation with the length of 20, the computational complexity of a distinguisher $\mathcal{D}$, executed by $\mathcal{S}$, in distinguishing the color values is 20! because $\mathcal{D}$ needs to figure out the correct one from 20! permutations, which means a $log_2(20!)$ $\approx$ 61 bits security strength. The information of $r$ is protected by block permutation and intra-block permutation. The security strengths of block permutation and intra-block permutation are equal to $log_2(blknum!)$ and $log_2(blksize!)$ bits, respectively. The $DC$ values is protected by the substitutions and bitxor by a random bit-stream. The security strengths of bitxor are equal to $n$, where $n$ means the length of random bit-stream. Above encrypted information compose the encrypted image, therefore, the security strength of image encryption $Sec_{Img}$ in our scheme can be calculated as:
	
	\begin{equation}
	\label{eq:secstrn}
	\begin{split}
	Sec_{Img}    = & 3 \times N_{pmt1} \times log_2(20!)\\
	+ & \sum log_2(blknum_*!) \\
	+ & 3 \times \sum\sum_{i=1}^{blknum_*}log_2(blksize_{i}!)\\
	+ & 3 \times N_{pmt2} \times log_2(10!)\\
	+ & \sum\sum_{i=1}^{blknum_*}n_{i}(bits)
	\end{split}
	\end{equation}
	
	\item \emph{Security of features}. In our scheme, image features are mainly calculated from the local histograms of encrypted $DC$ and $(r,v)$ pairs. With a simulated image $I^\mathcal{S}$, $\mathcal{S}$ can calculate the local histograms of the simulated image. The computational complexity of a distinguisher $\mathcal{D}$ in distinguishing the histogram is $3 \times N_{pmt1} \times log_2(20!) + 3 \times N_{pmt2} \times log_2(10!)$ which means about 642 bits security strength if we set $N_{pmt1} = N_{pmt2} = 5$
	
	\item \emph{Security of query image and its feature}. As shown in algorithm \ref{alg: TrapGen}, all the queries are encrypted by different $ImgPosKey$ and $ImgValKey$. The one-time pad encryption makes the multiple $\{\mathcal{C}\}$ the same security level with single $\mathcal{C}$. As the query is firstly encrypted like algorithm \ref{alg: OwnImgEnc}, which makes the security of the query not less than images in CS. What's more, the extra \textsf{BlockPerm} avoids the leakage of the relation between the encrypted blocks.
	
\end{itemize}

The images are encrypted by the combination of block permutation, intra-block permutation, value substitution, and bitxor in JES-MSIR. Although the security is partly depending on the image size, the efficiency advantage makes this kind of encryption more suitable for images compare to the methods based on HE \cite{hu2016securing}.

\subsubsection{Security under KBA model}

\begin{figure}[ht]
	\centering
	\includegraphics[width=1.0\linewidth]{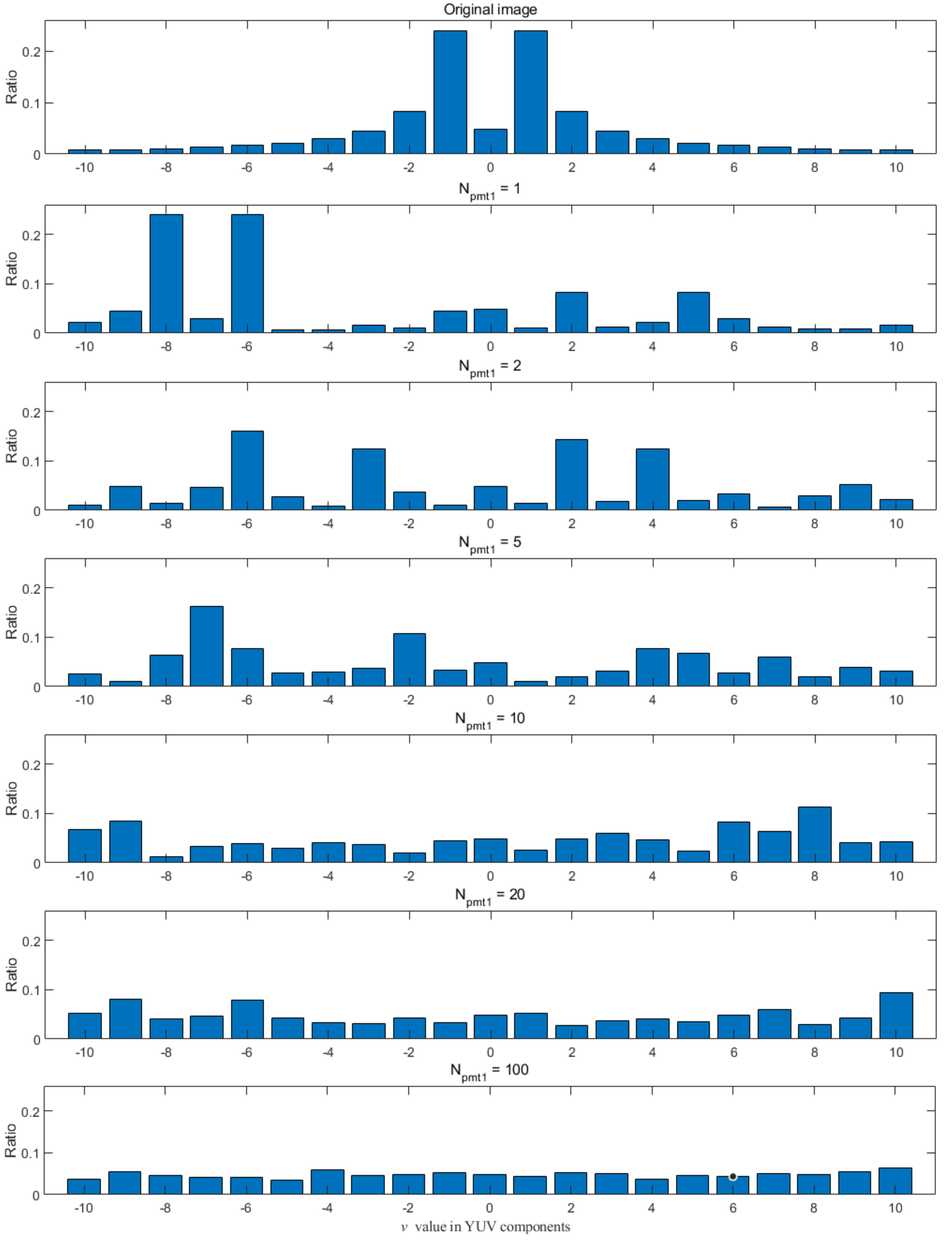}
	\caption{Occurrence ratios of $v$ values of original image and its encrypted versions with different $N_{pmt1}$. \cite{xia2019boew}}
	\label{fig:vfrequencyinKBA}
\end{figure}

In the KBA model, the adversary also knows certain statistical properties of natural images, which degrades the security strength of the proposed scheme. For instance, as shown in the first subfigure of Fig. \ref{fig:vfrequencyinKBA}, $v$ values do not occur uniformly, and the small $v$ value generally has a much higher frequency. After the substitution with a single permutation table, although the histogram bins have been shuffled, however, the distribution statistics are still reserved as shown in the second subfigure of Fig. \ref{fig:vfrequencyinKBA}. In this case, the CS which always has an image database in plaintext is easy to infer the secret permutation. In our scheme, multiple permutation tables are utilized, which will flatten the $v$ value histogram of the encrypted image and thus offering stronger security. Although the histogram becomes flatten, the size relationship of retrieval distance is approximately kept. It is clearly a trade-off between security and retrieval accuracy. In this paper, we set $N_{pmt1}$ = $N_{pmt2}$ = 5.

\subsection{Security under Collusion}\label{sec:collusion}
In the above analysis, we prove that our scheme is safe if each participant in the system is reliable. However, a reasonable system should be robust to the collusion between users. As the image stored in CS, we skip the analysis of collusion between KMC and member in the group or user.

\subsubsection{The collusion between CS and KMC}
Colluding CS and KMC own the knowledge of the encrypted images and encrypted key. The security of encrypted images has been shown in the COA model, here we further prove KMC can not infer the $ImgPosKey$ from encrypted keys.

As formula \ref{eq:decencenc} shows, to the \textsf{EncPerm} and \textsf{DecPerm}, there are $n!$ possible permutations for $n$ elements. It means if all $K$, $K_1$, $K_2$ are unknown, it is indistinguishable to infer them from the $\textsf{DecPerm}(K_2, K_1)$ and $\textsf{EncPerm}(K_1, K)$ only. Here the $ImgPosKey$ is unknown as shown in subsection \ref{sec:non-collusion}, $UserKey_{OID}$ are kept in the owner side, and $UserKey_{UID}$ are stored in the authorized user side. As both CS and KMC are unfamiliar to the above information, it is difficult for them to infer $ImgPosKey_{(OID,IID)}$ from $UserKey_{(OID,UID)}$ and $ImgPosKey_{(OID,IID)}^{'}$. In an ideal environment (i.e., non-collusion), the CS and KMC can be undertaken by one server.

\subsubsection{The collusion between CS and authorized users}

Colluding CS and AU know the decrypted images and $ImgValKey_{OID}$, which means the features are exposed. However, the images stored in CS are encrypted with the one-time-pad $ImgPosKey$, in this case, the colluder can't obtain the unknown images. As the image features are meaningless to authorized users, it will further decrease the possibility of collusion.

\subsubsection{The collusion between CS, KMC and a separate image owner}\label{subsec:colludeCSAndIO} 
The image owner may try to know more images from other image owners through colluding with the CS. As different owner use different $ImgValKey$, we here mainly consider the potential colluding risk on the AU who authorized by conspirator and other image owners.

As Algorithm \ref{alg: TrapGen} shows, each encrypted query is finally protected by \textsf{BlockPerm}, which means the conspirator can not get the block relation between the queries. Take $v$ for instance, as we use multi-table in the ValueSubstitution, which means each block may be encrypted by $N_{pmt2}$ possibilities. The owner can not infer the plaintext query feature in that there are $(N_{pmt2})^{blknum}$ possibilities for an image that has $blknum$ blocks. It further means the collusion with one image owner will not expose $ImgValKey$ of the image owners authorized to the same user. As KMC has no information related to $ImgValKey$, this kind of collusion will not leak the plaintext image.

\subsubsection{The collusion between CS and member in the group}\label{subsec:colludeCSandMember}

After collusion, CS has the knowledge of $ImgValKey_{GID}$ and $UserKey_{GID}$, which makes the image feature belong to the group exposed. However, as the $ImgPosKey$ is one-time-pad, it can not be directly exposed. Further, the key KMC sends to CS is encrypted by AES if it corresponding to a group, which means CS has no information related to $UserKey_{GID}$. Therefore, the position information of the image is still secure in this situation.

\subsection{The security comparison}\label{subsec:securitycompare}
Here we compare the security with the former paper from the perspective of image content and image feature, and the conclusion can be seen in table \ref{tab:securitycomparison}. In \cite{shen2018content}, to avoid the key conversion during image decryption, the KMC decrypts and gets the plaintext image, which makes their scheme insecure. In \cite{zhang2017pic}, the feature security depends on a global key, which can be got by colluding CS and KMC.

\begin{table*}[ht]
	\centering
	\caption{Security comparison}
	\label{tab:securitycomparison}
	\begin{tabular}{|l|c|c|c|c|c|l|c|}
		\hline
		\multicolumn{1}{|c|}{\multirow{2}{*}{}} &
		\multicolumn{2}{c|}{JES-MSIR} &
		\multicolumn{2}{c|}{MIPP \cite{shen2018content}} &
		\multicolumn{3}{c|}{PIC \cite{zhang2017pic}} \\ \cline{2-8} 
		\multicolumn{1}{|c|}{} &
		\multicolumn{1}{l|}{content} &
		\multicolumn{1}{l|}{feature} &
		\multicolumn{1}{l|}{content} &
		\multicolumn{1}{l|}{feature} &
		\multicolumn{2}{l|}{content} &
		\multicolumn{1}{l|}{feature} \\ \hline
		No Collusion           & yes & yes & no & no & \multicolumn{2}{c|}{yes} & yes \\ \hline
		Colluding CS and KMC   & yes & yes & no & no & \multicolumn{2}{c|}{yes} & no  \\ \hline
		Colluding CS and User  & yes & no  & no & no & \multicolumn{2}{c|}{yes} & yes \\ \hline
		Colluding CS, KMC and seperate Owner & yes & yes & no & no & \multicolumn{2}{c|}{yes} & no \\ \hline
		Colluding CS and member& yes & no  & -  & -  & \multicolumn{2}{c|}{-}   & -   \\ \hline
	\end{tabular}
\end{table*}

\section{Experiment results}\label{sec:section7}
The section evaluates the performance of the proposed scheme in terms of encryption effectiveness, retrieval accuracy, and retrieval efficiency. We implement the proposed scheme with Matlab 2018a on a Windows 10 operation system. All the experiments on the user side (i.e., source and authorized user) are executed in a machine with Intel Core i5-8250u CPU @ 1.6GHZ and 16GB memory. The experiment in the Cloud side (i.e., CS or KMC) is executed on a machine with Intel Core i7-6900K CPU @ 3.20GHz and 64 GB memory. We firstly use the commonly used Corel-1k image dataset \cite{li2003automatic} as the experiment dataset. The images in this dataset size either 384$\times$256 or 256$\times$384. The image dataset includes 10 categories and each category contains 100 similar images.

\subsection{Upload/Update consumption}

In this section, we focus on the time consumption in the image owner side. Generally speaking, in the existing schemes, the image owner needs to execute the following sub-operation: \emph{Image encryption, Feature extraction, Feature aggregation, Feature encryption}. The time consumption comparison on uploading Corel-1k dataset is shown in Table \ref{tab:imgupload}. Benefiting from the simple encryption scheme, the \cite{ferreira2017practical} and \cite{cheng2016encrypted} are in high efficiency. Suffering from the high computation complexity of FHE, the encryption on feature is also an expensive operation in \cite{zhang2017pic}.

\begin{table*}[htb]
	\centering
	\caption{Time consumption of image dataset upload}
	\label{tab:imgupload}
	\begin{tabular}{|l|c|c|c|c|c|}
		\hline
		& JES-MSIR  & {Cheng \cite{cheng2016encrypted}} & {IES-CBIR \cite{ferreira2017practical}} & {MIPP \cite{shen2018content}} & {PIC \cite{zhang2017pic}}   \\ \hline
		Image encryption         & 90.1s & 79.07s & 47.66s & 2.51s  & 2.7s    \\ \hline
		Feature extraction       & -     & -      & -      & 13s    & 77.43s   \\ \hline
		Feature aggregation           & -     & -      & -      & -      & 200.3s   \\ \hline
		Feature encryption & -     & -      & -      & 6.7s   & 1228.8s  \\ \hline
		Total time consumption   & 90.1s & 79.07s & 47.66s & 22.21s & 1509.23s \\ \hline
	\end{tabular}
\end{table*}

Further, the image owner has the need for updating their image. The existing schemes execute the following sub operations during the update: \emph{Image update, Feature update}. The time consumption in update is shown as Table \ref{tab:imgupdate}. The update operation in \cite{shen2018content,ferreira2017practical,cheng2016encrypted}, and JES-MSIR is similar with the uploading. In \cite{zhang2017pic}, the image owner only needs to generate the feature based on existing visual words. However, it still leads to a costly update.

\begin{table*}[h]
	\centering
	\caption{Time consumption of image update}
	\label{tab:imgupdate}
	\begin{tabular}{|l|c|c|c|c|c|}
		\hline
		& JES-MSIR  & {Cheng \cite{cheng2016encrypted}} & {IES-CBIR \cite{ferreira2017practical}} & {MIPP \cite{shen2018content}} & {PIC \cite{zhang2017pic}} \\ \hline
		Image update         & 0.09s & 0.08s  & 0.05s  & 0.03s  & 0.04s  \\ \hline
		Feature update       & -     & -      & -      & 0.02s  & 7.35s      \\ \hline
		Total time consumption & 0.09s & 0.08s  & 0.05s  & 0.05s  & 7.39s  \\ \hline
	\end{tabular}
\end{table*}

\begin{figure}[ht]
	\centering
	\includegraphics[width=1.0\linewidth]{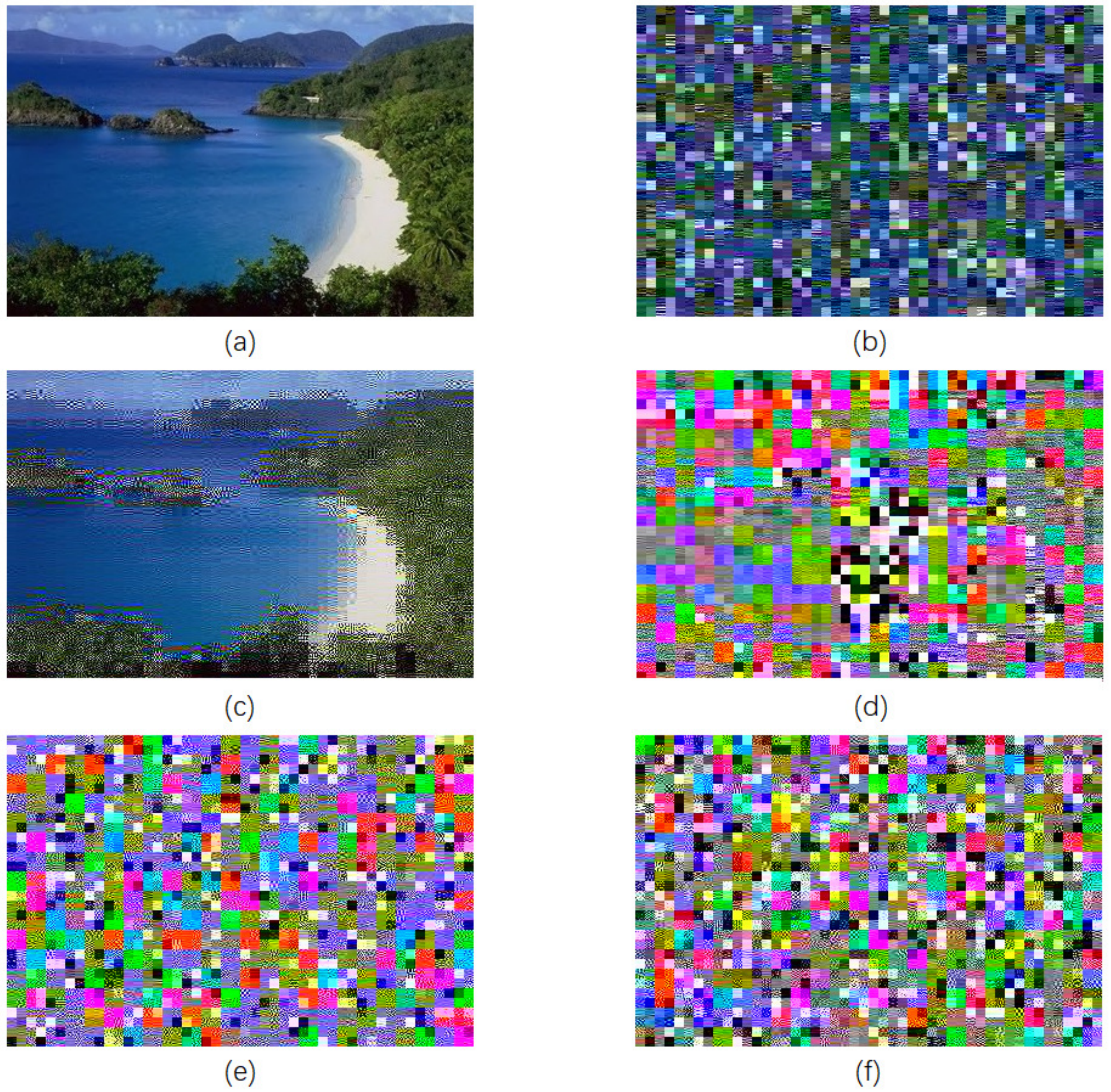}
	\caption{The visual effect of encryption, (a) the original image (133.jpg in Corel-1k database), the size of which is 384$\times$256, (b) with block permutation only, (c) with intra-block permutation only, (d) with value substitution only, (e) with value substitution under $N_{pmt1,pmt2} = 1$,  (f) with value substitution under $N_{pmt1,pmt2} = 5$.}
	\label{fig:encryptedImage}
\end{figure}

The time consumption of transferring the image to the CS and the following operation on the encrypted images is almost linear to the size of the encrypted image, we further give the size of encrypted image information in Table \ref{tab:sizeincreasement}. The size of encrypted image in \cite{shen2018content,zhang2017pic} are equal to plaintext. Fig. \ref{fig:encryptedImage} illustrates the separate and joint effect of the three protecting steps. It should be note that the multiple permutations lead to more uniform encrypted pixels.

\begin{table*}[ht]
	\centering
	\caption{The size increment of the encrypted image dataset}
	\label{tab:sizeincreasement}
	\setlength{\tabcolsep}{3.8mm}{
		\begin{tabular}{|l|c|c|c|c|}
			\hline
			& Plaintext & JES-MSIR   & {Cheng \cite{cheng2016encrypted}} & {IES-CBIR \cite{ferreira2017practical}} \\ \hline
			Corel-1k dataset  & 32.3MB    & 57.1MB & 40.5MB & 268MB  \\ \hline
		\end{tabular}
	}
\end{table*}

\subsection{Retrieval Consumption and Precision}

In our experiments, the "precision" for a query is defined as that in \cite{muller2001performance}: $P_m = m'/m$, where $m'$ is the number of real similar images in the $m$ retrieved images. We choose all 10 categories to test retrieval precision and time consumption.

\subsubsection{single-source}
During the retrieval, the time consumption is composed of three parts: \emph{Trapdoor generation, similarity consumption in cloud side, decryption}. The time consumption comparison on retrieval (return Top-50 similar images) is shown as Table \ref{tab:retrievaltime}. Due to the leakage of the index, the similarity computation in \cite{cheng2016encrypted} is unacceptable. The retrieval consumption in \cite{zhang2017pic} depends on the utilized codebook as their scheme only compare with images in the same index each time.

\begin{table*}[h]
	\centering
	\caption{Time consumption of retrieval(Top-50)}
	\label{tab:retrievaltime}
	\begin{tabular}{|l|c|c|c|c|c|}
		\hline
		& JES-MSIR  & {Cheng \cite{cheng2016encrypted}} & {IES-CBIR \cite{ferreira2017practical}} & {MIPP \cite{shen2018content}} & {PIC \cite{zhang2017pic}}             \\ \hline
		Trapdoor generation & 0.09s & 0.08s  & 0.05s  & 0.03s  & 15.36s             \\ \hline
		similarity computation in cloud           & 0.11s & 75.63s & 0.15s  & 2.63s  & \textgreater{}600s \\ \hline
		Decryption          & 4.43s & 3.92s  & 2.44s  & 1.28s  & 1.31s             \\ \hline
		Total time consumption & 4.63s & 79.63s & 2.64s & 3.94s & \textgreater{}616.67s \\ \hline
	\end{tabular}
\end{table*}

\begin{figure}[tb]
	\centering
	\includegraphics[width=1.0\linewidth]{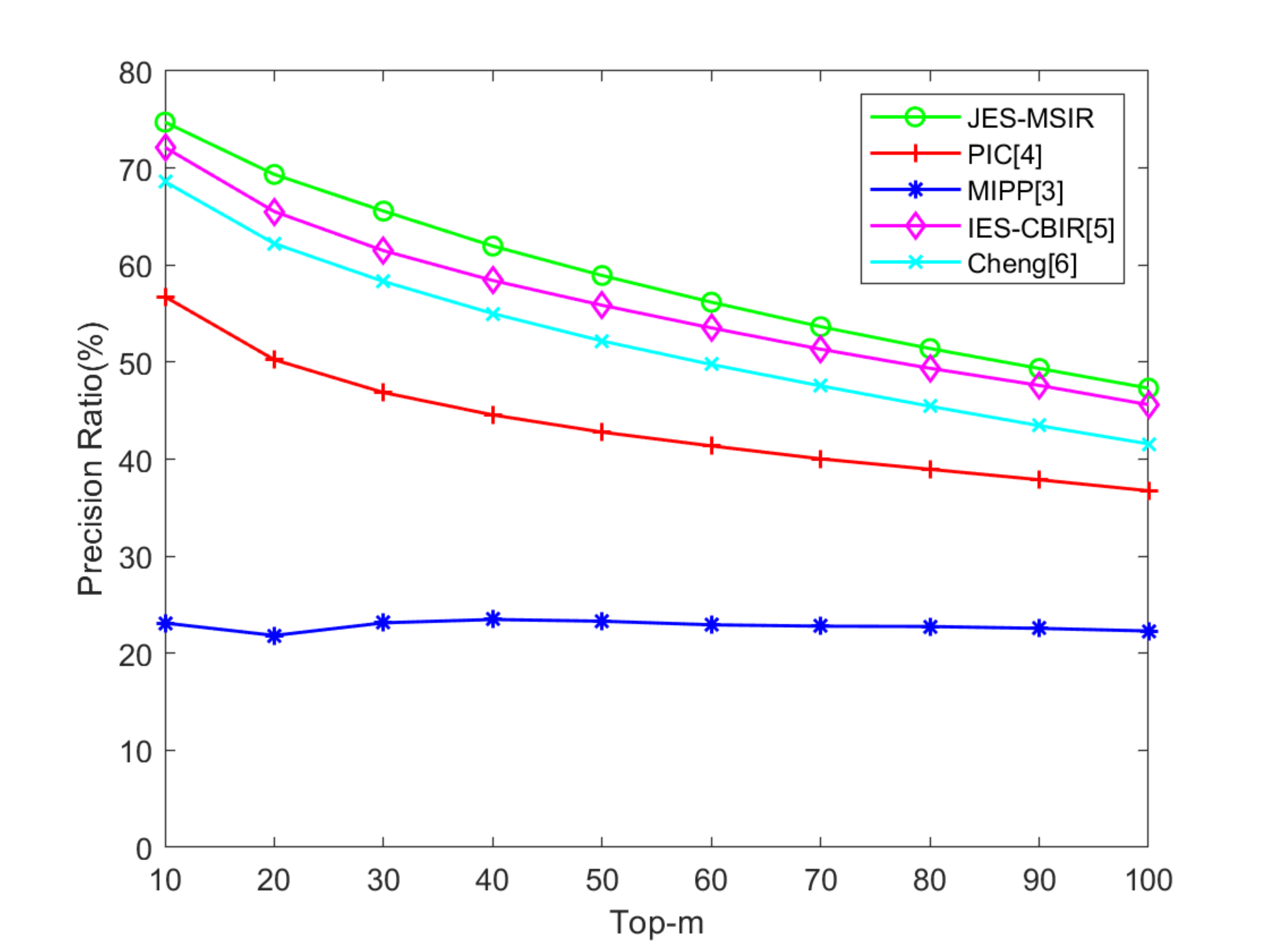}
	\caption{Retrieval accuracy comparison in Corel-1k dataset}
	\label{fig:retrievalAccuracyCorel1k}
\end{figure}

The retrieval accuracy comparison is shown as Fig. \ref{fig:retrievalAccuracyCorel1k}. Benefiting from the fully utilization of $DC$ and $(r,v)$ pairs and aggregation on local $(r,v)$ pairs, the accuracy of JES-MSIR is better than \cite{ferreira2017practical,cheng2016encrypted}. As the typical feature (e.g. SIFT) relies on effective aggregation schemes \cite{HolidayRes}, the accuracy in \cite{zhang2017pic} is not outstanding. However, more effective aggregation methods always lead to heavier computation consumption which will be undertaken by the image owner in the schemes belong to the first category.

\subsubsection{multi-source} \label{subsec:multi-source}
To better show the results in the multi-source scene, Corel-10k image dataset \cite{wang2001simplicity} is utilized. This image database includes 100 categories of images and each category contains 100 similar images. The size of images is either 187$\times$126 or 126$\times$187. We choose all 100 categories to test retrieval precision and time consumption. In our experiment, the images in Corel-10k dataset are randomly distributed to each image owner, and all the image owners possess the whole 10,000 images.

The retrieval time consumption is similar to the situation in single-source. Although our scheme needs to encrypt multiple queries, however, the time of trapdoor generation is far less than the other steps. Especially, the interaction rounds during the retrieval are shown in TABLE \ref{tab:Interactionrounds}. The interaction between CS and KMC in \cite{zhang2017pic} is unsure in that they can not ensure two rounds of interaction can get enough similar images.

\begin{table}[h]
	\centering
	\caption{Interaction rounds during the retrieval}
	\label{tab:Interactionrounds}
	\begin{tabular}{|l|c|c|c|}
		\hline
		& JES-MSIR & {MIPP \cite{shen2018content}} & {PIC \cite{zhang2017pic}}           \\ \hline
		CS and KMC   & 1    & 1      & $\geq$2 \\ \hline
		CS and User  & 1    & 1      & 1                \\ \hline
		KMC and User & 0    & 1      & 0                \\ \hline
	\end{tabular}
\end{table}

\begin{figure}[tb]
	\centering
	\includegraphics[width=1.0\linewidth]{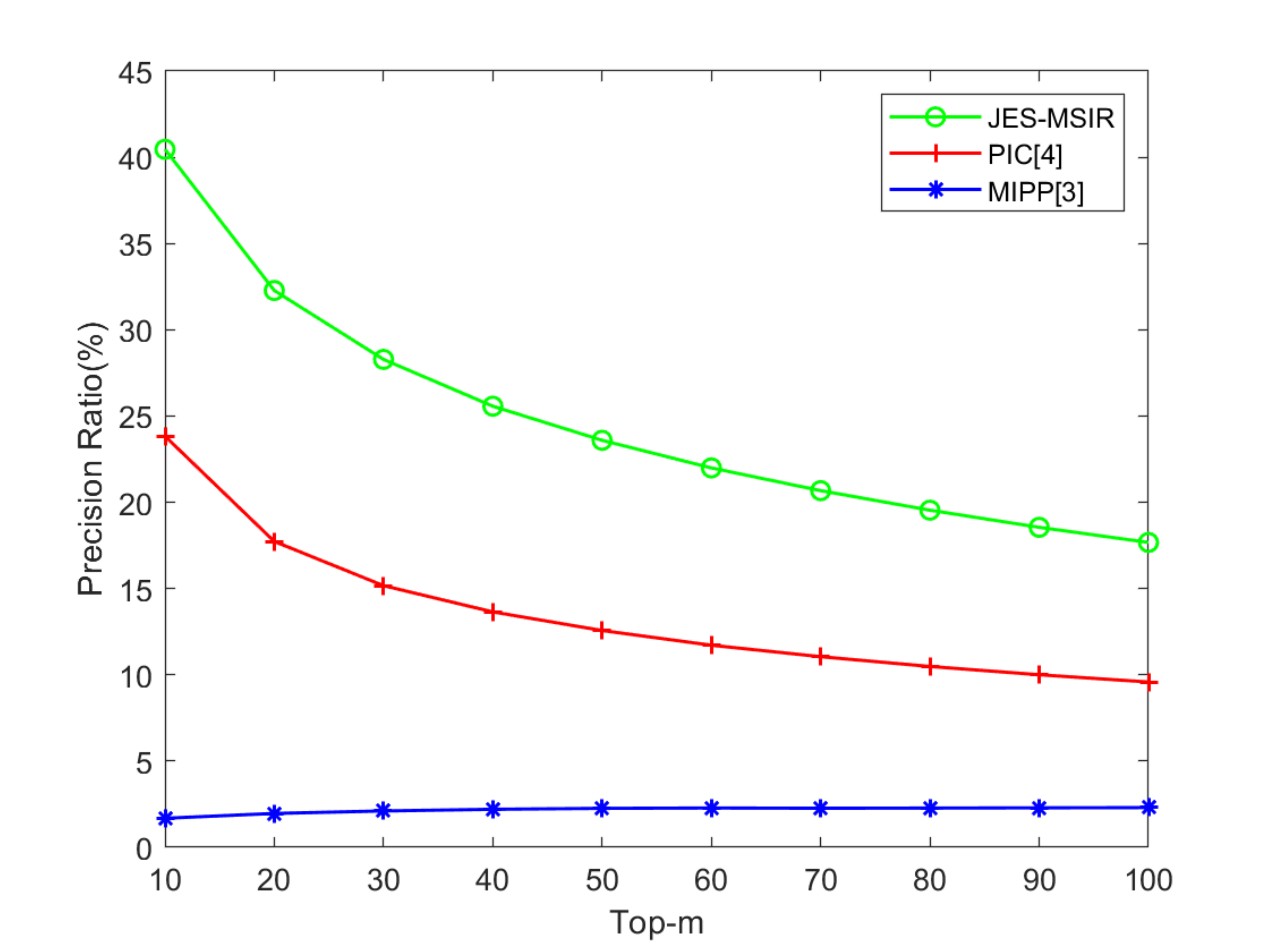}
	\caption{Retrieval accuracy comparison in Corel-10k dataset}
	\label{fig:retrievalAccuracyCorel10k}
\end{figure}

Fig. \ref{fig:retrievalAccuracyCorel10k} uses the Corel-10k database shows the retrieval accuracy comparison in the single-source scene. When $N_{source}$ (i.e., the number of source) increases, the accuracy of \cite{shen2018content} will be kept the same as the feature they use unchanged; the accuracy of \cite{zhang2017pic} will have an extent of change as the image owners jointly maintain the same codebook. When the images significantly increase, the quality of the codebook will infect the accuracy, the influence is basically same as the plaintext image retrieval \cite{chierichetti2007finding}. It should be noticed it is not robust as the alternation of codebook needs the participation of image owners.

In JES-MSIR, as mentioned in subsection \ref{sec:IdxForMulSource}, the distance are still in the same level if the same number of cluster centers are chosen. To choose reasonable $k_{g}$ for the system, we firstly use grid search to choose the approximately optimal $k_{grid}$ for the 1-source ($k_Y = 200$, $k_U = 50$, $k_V = 50$). Then we use the $k_{gird}$ as the $k_{g}$ to test the situation on different $N_{source}$. Further, the two, ten, half, tenth times of $k_{grid}$ are utilized as $k_{g}$ to test the robustness.

\begin{figure}[h]
	\centering
	\includegraphics[width=1.0\linewidth]{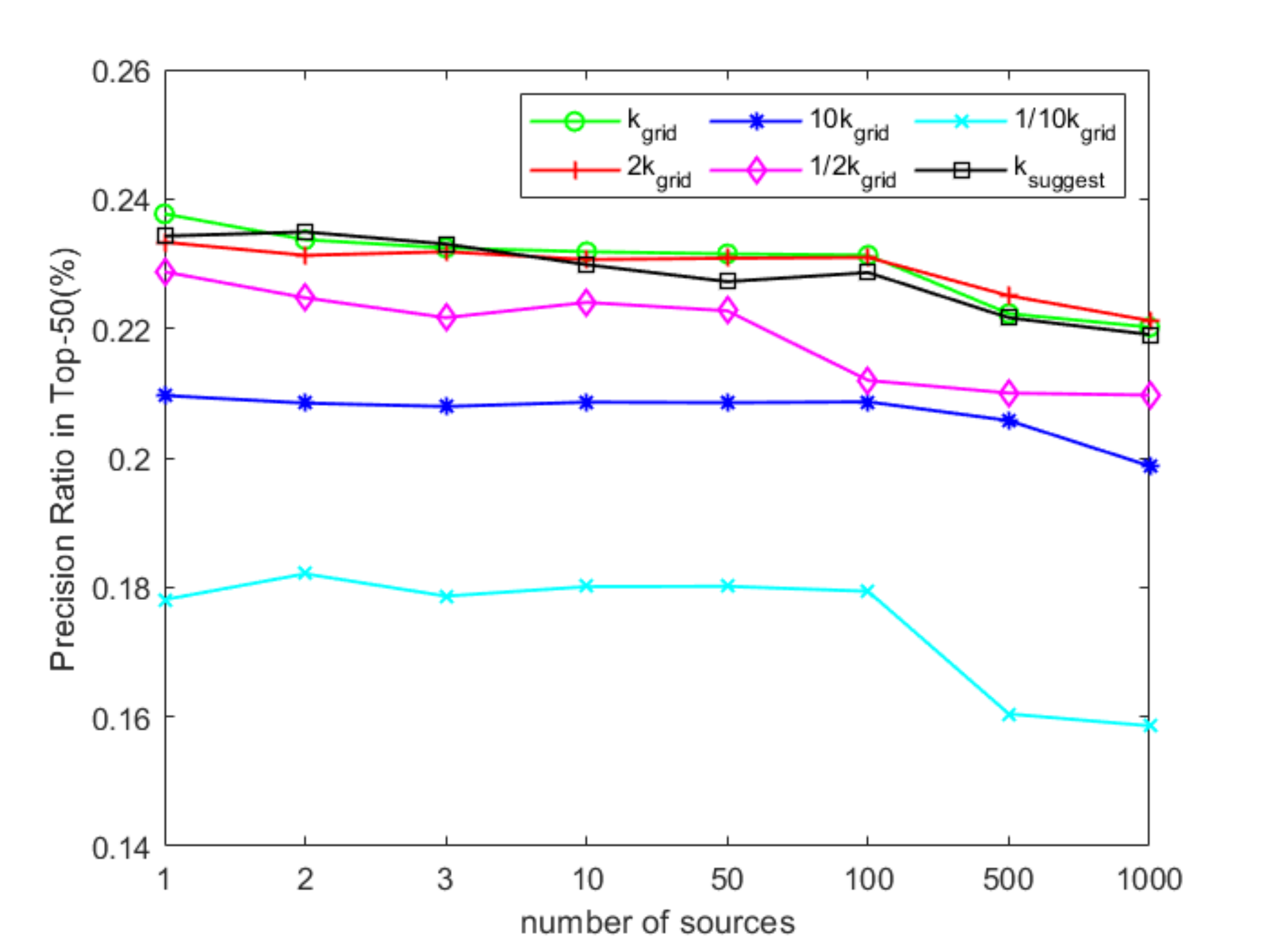}
	\caption{Top-50 accuracy comparison in different choice of $k$}
	\label{fig:DiffK}
\end{figure}

\begin{figure}[h]
	\centering
	\includegraphics[width=1.0\linewidth]{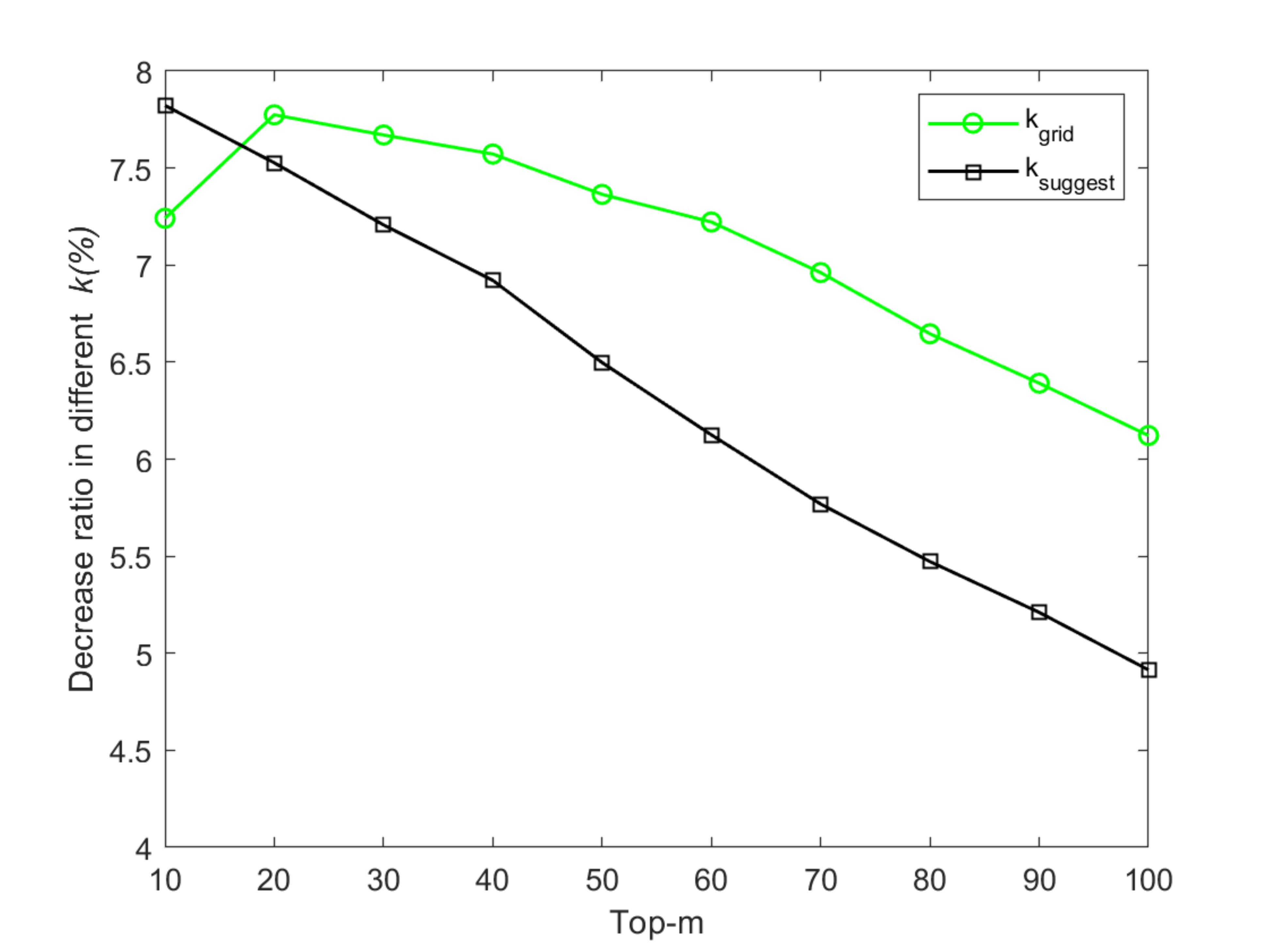}
	\caption{Retrieval accuracy decrease ratio in different Top-m}
	\label{fig:DecreaseRatio}
\end{figure}

As shown in Fig. \ref{fig:DiffK}, three conclusions can be seen. Firstly, the retrieval accuracy decrease at a slow speed in the same choice of cluster number. For instance, the retrieval accuracy (Top-50) only decrease 7.3\% when $N_{source}$ increase from 1 to 1,000. It means even in an extreme situation (i.e., each image owner has average 10 images), the retrieval accuracy is still stable. And the decrease ratio is in decline with the increment of returned images as shown in Fig. \ref{fig:DecreaseRatio}. Secondly, a small increment of $k$ is beneficial to accuracy. For instance, two times $k_{grid}$ gets better accuracy when $N_{source}$ over 500. Last but not least, the change in $k_{g}$ shows little influence on the results. It should be noticed that only 10\% accuracy loss when the 10 times $k_{grid}$ in utilization. And the accuracy is still better than \cite{zhang2017pic} even in the extreme situation (i.e., $k_Y$=20, $k_U$=5, $k_V$=5).

As most of methods which can infer the $k_{sug}$ need consume plenty of resources when the feature is huge. In this case, only part of feature are randomly chosen from the original feature as an optimization in our experiment, and the proportion of chosen feature can be briefly called $ratio$, where $ratio \in (0,1]$. Gap statics \cite{tibshirani2001estimating} method is employed in the experiment to get the $k_{sug}$. A sub-linear speed is gotten when the number of image decline. Consider all the above factors comprehensively, we here suggest the $k_{g}$ chosen as $k_{suggest}$ = $\frac{1}{N_{source}} (\sum_{i=1}^{N_{source}} k_{sug}^{i})\cdot log_2(1+\frac{1}{ratio}) \cdot log_2(1+N_{source})$. The result is shown in Fig. \ref{fig:DiffK} and Fig. \ref{fig:DecreaseRatio}, it could be noted that appropriate accuracy and slower decline can be got in different $N_{source}$. What's more, as all the feature aggregation tasks are undertaken by CS, it is easy for CS to update the $k_{g}$ at regular intervals.

\section{Conclusion}\label{sec:section8}
In this paper, we introduce the MSPPIR problem and propose a novel scheme that can effectively and securely cope with this problem. Different from the previous schemes which use the homomorphic encryption, we propose a scheme based on the randomization encryption, which leads to better efficiency, accuracy, and security. The bitxor and permutation are used to ensure the security of the image, and the BOW model is used to aggregate the encrypted $(r,v)$ pairs in a multi-source scene. As the retrieval accuracy is still insufficient when compared with that in the plaintext domain, in the future, we consider executing the state-of-art CBIR scheme in safety based on two non-collusion CS.

\section*{Acknowledgements}
This work is supported in part by the National Natural Science Foundation of China under grant numbers 61672294, 61502242, 61702276, U1536206, U1405254, 61772283, 616 02253, 61601236, and 61572258, in part by Six peak talent project of Jiangsu Province (R2016L13), in part by the Priority Academic Program Development of Jiangsu Higher Education Institutions (PAPD) fund, in part by NRF-2016R1D1A 1B03933294, in part by the Jiangsu Basic Research Programs-Natural Science Foundation under grant numbers BK2015092 5 and BK20151530, in part by the Collaborative Innovation Center of Atmospheric Environment and Equipment Technology (CICAEET) fund, China. Zhihua Xia is supported by BK21+ program from the Ministry of Education of Korea.

\bibliography{MSPPIR}

\bio{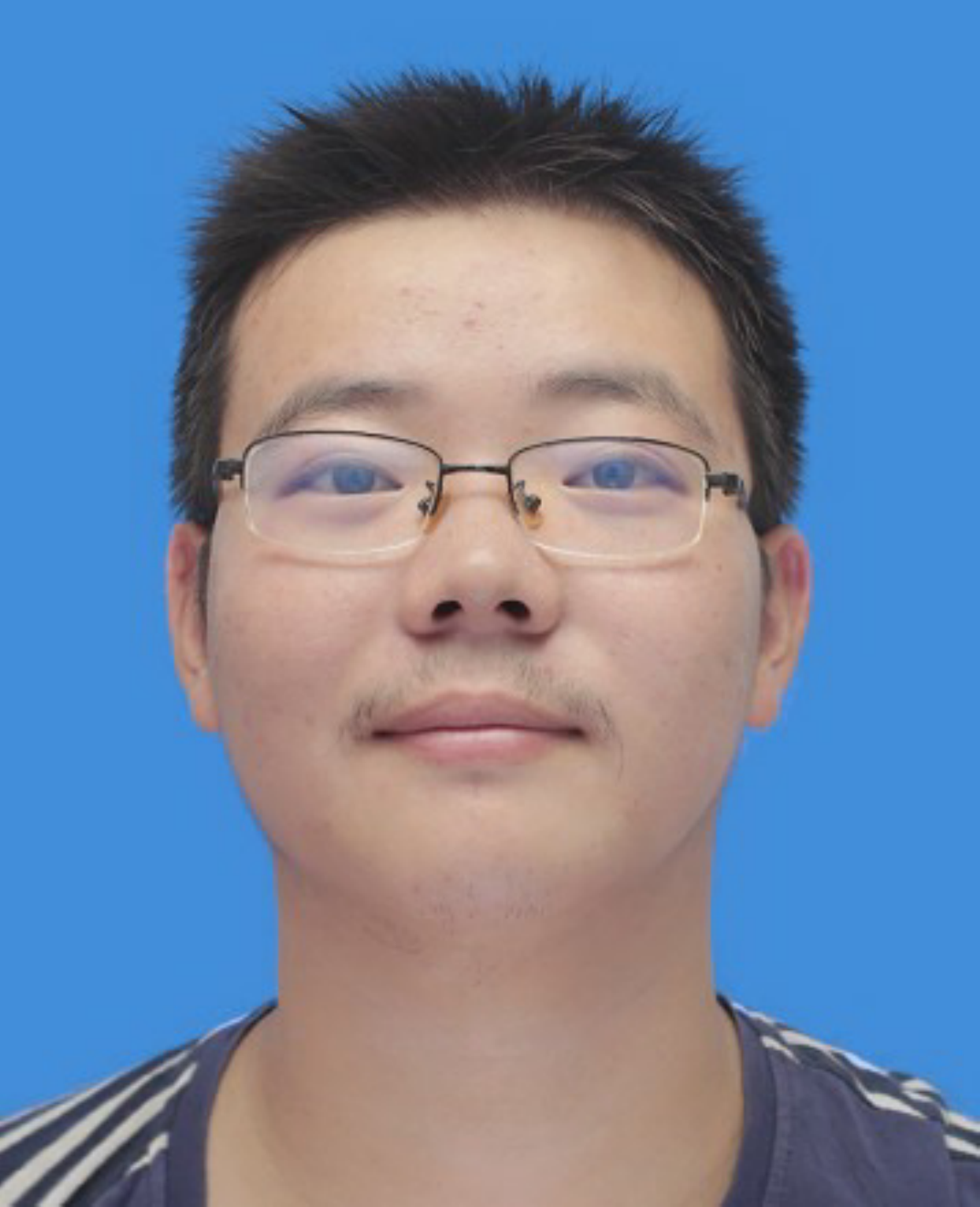}
Qi Gu is currently pursuing his master degree in the School of Computer and Software, Nanjing University of Information Science and Technology, China. His research interests include functional encryption, image retrieval and nearest neighbor search.
\endbio

\vskip60pt

\bio{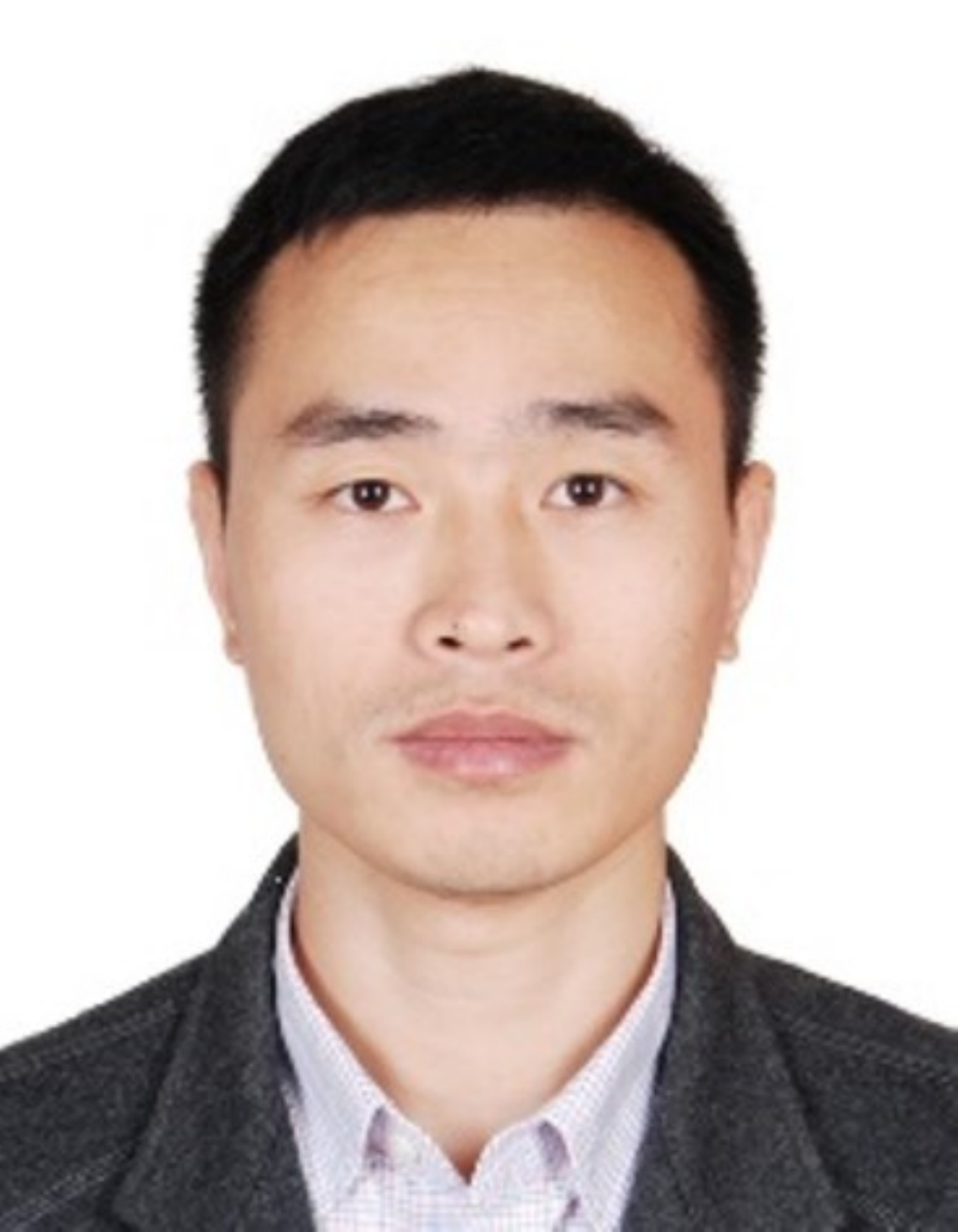}
Zhihua Xia received a BS degree in Hunan City University, China and PhD degree in computer science and technology from Hunan University, China, in 2006 and 2011, respectively. He works as an associate professor in the School of Computer and Software, Nanjing University of Information Science and Technology. His research interests include digital forensic and encrypted image processing. He is a member of the IEEE from 1 March 2014.
\endbio

\vskip20pt

\bio{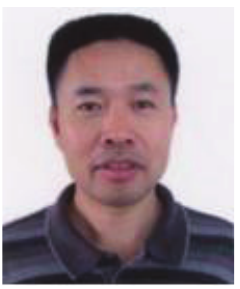}
Xingming Sun received his BS in mathematics from Hunan Normal University, China, in 1984, MS in computing science from Dalian University of Science and Technology, China, in 1988, and PhD in computing science from Fudan University, China, in 2001. He is currently a professor in China-USA Computer Research Center, China. His research interests include network and information security, digital watermarking, and data security in cloud.
\endbio

\end{document}